\documentclass{article}
\PassOptionsToPackage{numbers,compress,sort}{natbib}

\usepackage[final]{neurips_2025}

\usepackage[utf8]{inputenc} 
\usepackage[T1]{fontenc}    
\usepackage[colorlinks=true, allcolors=blue]{hyperref}       
\usepackage{url}            
\usepackage{booktabs}       
\usepackage{amsfonts}       
\usepackage{nicefrac}       
\usepackage{microtype}      
\usepackage{xcolor}         
\usepackage{placeins}
\usepackage[ruled,vlined]{algorithm2e}
\usepackage{float} 

\usepackage{amsmath}
\usepackage{graphicx}
\usepackage{subcaption}
\usepackage{svg}
\usepackage{wrapfig}
\usepackage{longtable}
\usepackage{threeparttable}
\usepackage{aliascnt}

\usepackage[capitalise,noabbrev]{cleveref}
\creflabelformat{equation}{#2\textup{#1}#3}

\title{Brain-Like Processing Pathways Form in Models With Heterogeneous Experts}

\author{%
 Jack Cook$^{1}$ \quad Danyal Akarca$^{2}$ \quad Rui Ponte Costa$^{1,*}$ \quad Jascha Achterberg$^{1,*}$\\
 $^1$Centre for Neural Circuits and Behaviour, University of Oxford\\
 $^2$Department of Electrical and Electronic Engineering, Imperial College London\\
 $^*$Joint senior authors}

\begin{document}
\maketitle

\begin{abstract}
    The brain is made up of a vast set of heterogeneous regions that dynamically organize into pathways as a function of task demands.
Examples of such pathways can be found in the interactions between cortical and subcortical networks during learning, or in sub-networks specializing for task characteristics such as difficulty or modality.
Despite the large role these pathways play in cognition, the mechanisms through which brain regions organize into pathways remain unclear.
In this work, we use an extension of the Heterogeneous Mixture-of-Experts architecture to show that heterogeneous regions do not form processing pathways by themselves, implying that the brain likely implements specific constraints which result in the reliable formation of pathways.
We identify three biologically relevant inductive biases that encourage pathway formation: a routing cost imposed on the use of more complex regions, a scaling factor that reduces this cost when task performance is low, and randomized expert dropout.
When comparing our resulting \textit{Mixture-of-Pathways} model with the brain, we observe that the artificial pathways in our model match how the brain uses cortical and subcortical systems to learn and solve tasks of varying difficulty.
In summary, we introduce a novel framework for investigating how the brain forms task-specific pathways through inductive biases, and the effects these biases have on the behavior of Mixture-of-Experts models.
\end{abstract}

\section{Introduction}

The brain is made up of many heterogeneous regions distinguished by features such as connectivity, cell types, neurotransmitters, and functional specialization \cite{yao2023high, glasser2016multi, shipp2007structure, mione2025neural}. To support complex behavior, the mammalian brain dynamically organizes these regions into diverse networks and processing pathways \cite{bertolero2015modular}, allowing it to adapt to different inputs and task demands. This principle spans sensory systems \cite{grill-spector_human_2004, de2011usefulness, arnott2004assessing}, cognitive networks \cite{duncan_construction_2025}, emotion-related circuits \cite{etkin_emotional_2011}, and face perception \cite{bernstein2015two}. Notably, pathway formation is highly dynamic: regions can participate in many pathways, allowing cognitive processes to arise from the joint activations of specific groups of regions. While theoretical work has shown how heterogeneous regions and modules can develop within networks \cite{yang2019task, achterberg_spatially_2023, sporns2016modular}, how these organize into large-scale pathways remains poorly understood.

The importance of studying pathway formation and coordination extends beyond neuroscience: it is also becoming increasingly relevant in machine learning research. As models have evolved from small networks with a couple of layers to large system-level architectures, achieving complex function while maintaining efficiency has become critical. One recent development toward this goal is the Mixture-of-Experts (MoE) architecture \cite{liu2024deepseek, jiang_mixtral_2024}, which contains specialized experts that are selectively activated based on the current input. This should create pathways between experts that selectively respond to inputs of varying complexity \cite{raposo2024mixture} to make efficient use of computational resources \cite{fedus2022review, barham2022pathways}.
However, this theorized specialization of experts appears to be limited in practice~\cite{gritsch2024nexus, olson2025semantic}, making it difficult for specialized task-complexity-related pathways to form.

These findings raise the question, how do stable and functionally relevant pathways form in networks of distributed heterogeneous experts? Do heterogeneous regions automatically group into such pathways, or are additional priors required? Once pathways develop in models, do they show the same context-aware adaptability that has been observed in the brain? To address these questions, we introduce a neural network architecture made up of heterogeneous experts to study the conditions under which processing pathways form, and the degree to which these pathways resemble those studied in the brain. Specifically, we adapt the Heterogeneous Mixture-of-Experts (HMoE) architecture~\cite{wang2024hmoe, jawahar2022automoe}, in which information may be dynamically routed to computational experts, or regions, of varying sizes. In our model, unlike prior work, each expert is implemented as a recurrent network that could be considered a standalone model or brain region. We study the pathway formation in this architecture while we train models to learn 82 time-series-based cognitive tasks of varying difficulty~\cite{khona_winning_2023}. Through these analyses, we find:
\begin{itemize}
    \item Layers of heterogeneous experts do not automatically form recognizable pathways.
    \item Instead, inductive biases are required for pathways to form: (i) a routing cost that penalizes the model for using larger experts, (ii) scaling the routing cost based on task performance, and (iii) random expert dropout. These result in the formation of a \textit{Mixture-of-Pathways}.
    \item The pathways that form in our new \textit{Mixture-of-Pathways} architecture mirror the interactions between cortical and subcortical pathways in the brain during learning, and are in line with the dynamics of the brain's multiple-demand system~\cite{duncan_multiple-demand_2010, duncan_construction_2025}.
\end{itemize}

To arrive at these findings, in the following we start by analyzing the usage of experts of a baseline model with HMoE layers.
We then develop specific inductive biases that encourage pathways to form, before finally comparing these pathways to observations made in the brain.
\section{Related Work}

Brain-like modularity and regional heterogeneity can be induced in neural networks through priors and training procedures to explain how such features develop in the brain \cite{yang2021towards, achterberg2023building, sporns2016modular, perez2021neural, bena2025dynamics, sheeran2024spatial}. The priors that are especially relevant in the context of this work relate to metabolic cost and energy efficiency, which are crucial in determining the brain's circuitry and function \cite{stroud2025effects, achterberg_spatially_2023, akarca2023weighted, gershman2015computational, ali2022predictive}.

While the above work often focuses on starting from fully-connected networks to observe the formation of modules and regions, modeling multi-region interactions has also become possible with modern methods \cite{mizes2024role, zhou2021distributed, pemberton2024cerebellar, fang2023predictive, moskovitz2024understanding, liebana2023striatal}. This line of work has revealed how joint computation can be implemented through interactions between independent modules \cite{bertolero2015modular}. However, these multi-region models generally predefine a specific circuit structure with a small set of regions, preventing further study on how regions come to interact in the first place. Notable work that allows for dynamic (non-fixed) interaction of multiple independent regions often assumes networks which are not able to learn tasks \cite{clark2025structure, pereira2024cognitive}, though \cite{kozachkov2022rnns} stands out with a trainable network made up of individual RNNs. Their multi-region networks can change their connectivity during learning, but cannot route information based on task context. The work most closely aligned with our goal is \cite{finzi2023single}, which studies how spatial (metabolic) constraints in feed-forward networks can form visual processing streams. However, it too does not consider how regions may change their interaction as a function of context, and does not allow for solving standard time-series-based cognitive tasks.

In the context of artificial intelligence, the introduction outlined how the popular Mixture-of-Experts architecture \cite{liu2024deepseek, jiang_mixtral_2024} is relevant to our question of how specialized regions dynamically organize into processing pathways. Recent efforts to build networks out of experts that vary in terms of their architecture \cite{wang2024hmoe, jawahar2022automoe, raposo2024mixture} and function \cite{gritsch2024nexus} are especially relevant here. Moreover, work has argued that routing pathways are a powerful method for handling the complex data-flow of these otherwise efficient architectures \cite{he2024expertflow, barham2022pathways}, but without investigating how adaptable pathways can be encouraged to form. In neuromorphic computing it has been shown to be possible to implement brain-like visual processing pathways to achieve efficient processing~\cite{yang2024vision}, but with a predefined static architecture.

\section{Methods}
\label{sec:methods}

In this work, we aim to identify the mechanisms by which pathways form between heterogeneous regions, and how those pathways are used across a diverse set of tasks. To study this computationally, we need a model made up of heterogeneous experts, analogous to brain regions, that work together to solve many different tasks. We create such a model by extending the Heterogeneous Mixture-of-Experts architecture \cite{wang2024hmoe} and training it on the Mod-Cog set of time-series-based cognitive tasks~\cite{khona_winning_2023}.

\subsection{Model Architecture: Heterogeneous Mixture-of-Experts}
\label{sec:hmoe}
\label{sec:model-architecture}

\begin{wrapfigure}{R}{0.50\textwidth}
    \vspace{-5pt} 
    \centering
    \includegraphics[width=\linewidth]{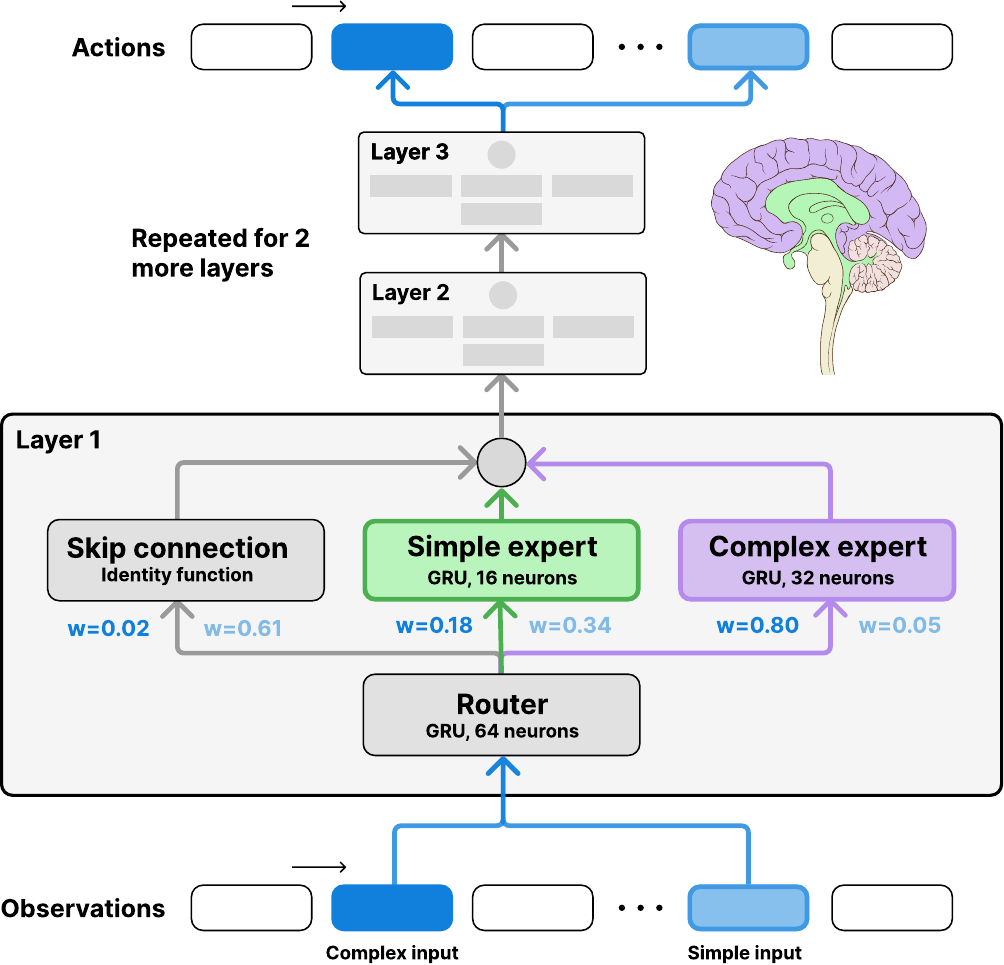}
    \caption{\textbf{Schematic of our baseline model architecture}. Information is passed through three layers, each of which can dynamically route information to experts of different computational complexity.}
    \label{fig:model-architecture}
\end{wrapfigure}

Mixture-of-Experts models (MoEs) \cite{liu2024deepseek, jiang_mixtral_2024} are characterized by their layers, which contain multiple smaller models alongside a router model, which decides which experts should process the input at each timestep. Specifically, the router determines the weight with which each expert contributes to the layer's final output. Experts can also be excluded, by setting an expert's weight to zero. In most modern MoEs, the experts are large scale MLPs placed in between attention layers, which are typically activated at every timestep. In Heterogeneous Mixture-of-Expert models \cite{wang2024hmoe} (HMoEs), the experts can vary in terms of their sizes and activation functions. We extend the HMoE architecture with several significant adaptations. Namely, each layer of our model contains three experts: two GRUs with 16 and 32 neurons respectively, and one skip connection, which allows the model to choose to perform no computation for a given timestep~\cite{raposo2024mixture}. Our implementation uses GRUs with 64 neurons as routers and does not include any additional layers between HMoE layers. We use this setup as a baseline for our investigations (\cref{fig:model-architecture}). In the following sections, we will introduce additional inductive biases to this baseline architecture, resulting in our final Mixture-of-Pathways model. The algorithm for this full model is described in \cref{algo:abbreviated}.

\begin{algorithm}
\SetAlgoLined
\KwData{Task dataset $D$, Experts $E = \{e_1, e_2, \ldots, e_n\}$, Routers $R = \{r_1, r_2, \ldots, r_n\}$}
Initialize routers and heterogeneous experts, set $h_0$ to task input\;
\For{each training step}{
    Sample batch $b$ from $D$\;
    \For{each task $i$, timestep $t \in b$}{
        \For{each layer $l$ with experts $e_j \in l$}{
            Compute routing weights $w_j = \text{softmax}(r_l(h_{l-1}))$ \;
            Apply expert dropout (Not in baseline architecture; see \cref{sec:expert-dropout}) \;
            Compute expert activations $w_j$ for each expert $e_j \in l$ \;
            Combine outputs: $h_l = \sum_j w_{j}h_{l,j}$\;
        }
        Compute baseline model losses: $L_{\text{fix}}$ and $L_{\text{response},i}$ \;
        Compute $L_{\text{routing}}$ loss (Not in baseline architecture; see \cref{sec:routing-stability}) \;
        Compute $L_{\text{total}} = L_{\text{fix}} + L_{\text{routing}} + \sum_i L_{\text{response},i}$\;
    }
    Update parameters using Schedule-Free AdamW optimizer~\cite{defazio2024road}\;
}
\caption{Mixture-of-Pathways training protocol. Full details in \cref{app:implementation-details}.}
\label{algo:abbreviated}
\end{algorithm}

\subsection{Evaluation with Cognitive Tasks}
\label{sec:methods-training}

To evaluate how pathways are formed and used across tasks with different characteristics, we use the Mod-Cog task set, which contains 82 time-series-based cognitive tasks \cite{khona_winning_2023}. This is an expansion of the popular NeuroGym framework \cite{molano2022neurogym}, which contains tasks like Go-NoGo or two-stimuli integration tasks (see \cref{app:task-visualisation} for task details). Importantly for us, the tasks vary in difficulty due to their varied inputs, decision rules, and delay lengths. Generally, tasks range from 0.8 to 3 seconds in duration, during which we sample information at timesteps 100 milliseconds apart. At each timestep, the model receives a 115-dimensional input made up of four components: a 1-dimensional fixation input, two 16-dimensional stimuli, and an 82-dimensional one-hot encoding of the active task, which we pass through a 16-dimensional learned embedding layer. While the fixation input is active, the model should always output zero. After the fixation period, the model needs to use the observed information to output the correct choices during the response period.

We train our models over 10 epochs, each containing 1000 training steps.
At each training step, models are given a $128 \times 350 \times 115$ matrix of input data, representing 128 batches of task sequences that are 350 timesteps long, with 115 features at each timestep.
These task sequences contain many individual tasks: the average task is about 20 timesteps long, meaning that in each batch, models observe about 27 trials of each task.
Models are trained with a cross entropy loss $L_{\text{response},i}$ for the correct response during the response period of task $i$. An additional fixation loss $L_\text{fix}$ encourages the model to output zero during the fixation period.
All losses used in our analyses are detailed in \cref{app:loss-overview}.
Training one model takes roughly 1 hour on a single NVIDIA T4 GPU.
Details on implementation and code access are outlined in \cref{app:implementation-details}.

Once models have learned to solve each task by routing information between experts, we can study the conditions under which processing pathways form between layers. In the following we will first study the routing behavior of our baseline architecture.
We will then show how the additional inductive biases described in \cref{algo:abbreviated} result in the formation of pathways.
Finally, we will test the degree to which these pathways resemble established processing pathways in the brain.
\section{What Causes Pathways to Form?}
\label{sec:pathway_training}
\label{sec:pathway_formation}

In this section, we investigate the conditions under which pathways form between layers of heterogeneous experts. We set three criteria to determine whether pathways have formed:

\begin{enumerate}
    \item Pathways should be \textbf{consistent} with respect to tasks, meaning that when two models are trained to solve the same tasks, they should have structurally similar pathways.
    \item Pathways should be \textbf{self-sufficient}, meaning that when experts outside of a pathway are removed, then the model's overall performance should remain largely intact.
    \item Pathways should be \textbf{distinct}, meaning that several different pathways should be used to solve groups of tasks with varying characteristics.
\end{enumerate}

\subsection{Pathway Consistency}
\label{sec:routing-stability}
\label{sec:loss-normalization}

To see if experts form consistent, task-driven pathways, we train 20 randomly initialized models with the same settings and compare their routing patterns on the same set of 82 cognitive tasks. This allows us to examine whether models use similar sets of experts to solve the same tasks, such as whether smaller experts are reliably used to solve simpler tasks, or vice versa. We first do this with a baseline model made up of three HMoE layers, and then test each model on 50 trials of each task while recording the routing weights assigned to each expert at each timestep ($w$ values in \cref{fig:model-architecture}). To test whether routing is stable across training runs, we use these weights to calculate each model's \textit{Learned Pathway Complexity} for each task $i$ ($LPC_i$) as follows:

\begin{equation}
    LPC_i=\frac{1}{T_i}\sum_t^{T_i}\sum_j^E w_{i,j,t}s_j^2
    \label{eqn:LPC}
\end{equation}

\begin{wrapfigure}{r}{0.6\textwidth}
    \vspace{-10pt}
    \centering
    \includegraphics[width=\linewidth]{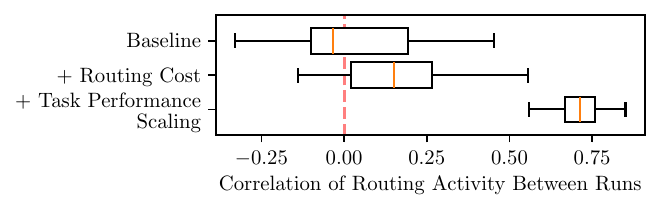}
    \caption{\textbf{Models trained with routing cost and task-based scaling exhibit more stable routing.} Correlations are calculated between the routing patterns across 20 training runs of each model setup.}
    \label{fig:routing-stability}
    \vspace{-0pt}
\end{wrapfigure}

This metric is calculated by multiplying the weight $w_{i,j,t}$ assigned by the router to each expert $j$ at each timestep $t$ by the squared size $s_j^2$ of expert $j$ while the model solves task $i$. This is then averaged across the total timesteps $T_i$ of each task $i$ to ensure that longer tasks are not biased toward having larger $LPC$s.
This results in a $LPC$ value for each of the 82 tasks and 20 model runs (see \cref{app:penalty_calc} for an example calculation).
The squaring of expert sizes is motivated by the $O(s_j^2)$ cost of storing each expert's weight matrix in memory, and we expand on the suitability of using $s_j$ as a measure of each expert's complexity in \cref{sec:effective-rank}.
Skip connections are free to use. To measure pathway consistency, we can now correlate this list of $LPC$ values across training runs. For the baseline model, we find that models are not consistent across training runs (mean pairwise correlation of 0.0324, \cref{fig:routing-stability}), suggesting that the baseline model does not form any stable and task-related processing pathways by default. Therefore, we next want to explore which specific inductive biases may result in such pathways.

Theories of metabolic optimization and cost minimization are core parts of our understanding of the brain's computations \cite{kool_mental_2018, bullmore2012economy, gershman2015computational}.
The reduction of energy consumption has been a powerful source of priors for building brain-like neural networks \cite{achterberg_spatially_2023, ali2022predictive, stroud2025effects} and more generally achieving brain-inspired computing \cite{yang2024vision, aimone2021roadmap}.
Hence we hypothesize that regularizing the routing weights by making it more expensive to route to more complex experts might cause replicable pathways to develop, as observed in the brain.
We do so by incorporating the $LPC_i$ (from \cref{eqn:LPC}) into the model's loss, making it more costly for the model to activate more complex experts.
Finally, to avoid convergence on the local minimum of only using the smallest experts without solving any tasks\footnote{Note that routers converging to local minima is an established phenomenon in Mixture-of-Experts models, as there is a bias to rely on the expert that learns the task, or decreases the overall loss as in our case, first~\cite{jiang_mixtral_2024,fedus_switch_2022,rajbhandari_deepspeed-moe_2022,wang2024hmoe}.}, we add a normalization strategy, dividing each $LPC_i$ by $L_{\text{response},i}$, the cross-entropy loss measuring the model's performance on task $i \in \mathcal{T}$.
In addition to helping with convergence, this normalization term can also be viewed as helping our model more directly control the cognitive effort, or processing power, with which it solves a task.
We discuss this further in \cref{sec:discussion}.

Adding these additional components to the loss results in the following equation, where $L_\text{fix}$ and $L_{\text{response},i}$ are the standard task-based losses described in \cref{sec:methods-training}.
A small value $\epsilon$ is added to ensure that if the model solves tasks perfectly, the routing loss is not $\infty$. All loss calculations are outlined in detail in Appendix \ref{app:loss-overview} and \ref{app:penalty_calc}.

\begin{equation}
L=L_\text{fix}+\sum_i^\mathcal{T}(L_{\text{response},i}+\frac{\alpha LPC_i}{L_{\text{response},i}+\epsilon})
    \label{eqn:cost-based-loss}
\end{equation}


We now evaluate the routing consistency of models trained with this custom loss function. Figure \ref{fig:routing-stability} shows that our expectations are confirmed: on its own, adding the $LPC_i$ for each task creates more consistent routing (mean pairwise correlation of 0.15, significant over baseline with $p<0.01$). Scaling this term by the model's performance on each task $L_{\text{response},i}$ amplifies this effect, encouraging models across training runs to converge on more consistent routing patterns (mean pairwise correlation of 0.71, significant over baseline with $p<0.001$).

\subsection{Self-Sufficiency of Pathways}
\label{sec:pathway-causality}
\label{sec:pathway-sufficiency}
\label{sec:expert-dropout}

Our second criterion measures whether formed pathways are self-sufficient, meaning that removing an expert that is not part of the currently activated pathway should only minimally impact the performance of the model. To test for self-sufficiency, we first evaluate whether models are still able to perform tasks when they are prevented from using experts that have been assigned low routing weights. We find that our baseline models are extremely sensitive to this deactivation: if they are prevented from using experts with $w<0.025$, which only contribute 2.5\% or less to each layer's output, average task accuracy drops from 98.2\% to 16.4\%. This shows that while models learn replicable routing patterns, these are not yet pathways, as they rely on all the experts.

We speculate that pathway self-sufficiency can be achieved by stochastic dropout of experts with low routing weights. Dropout is especially interesting as it is an established principle for achieving more robust neural networks \cite{srivastava_dropout_2014} and has also been linked to the stochastic nature of signal processing in neuroscience \cite{deco_stochastic_2009, li2021adaptive}. We implement \emph{expert dropout} by randomly deactivating experts that contribute very little to the output during training. The probability $p_j$ with which expert $j$ is deactivated at a given timestep is determined as follows:

\begin{equation}
    p_j= 
\begin{cases}
    \beta-\frac{\beta}{\gamma}w_j,& \text{if } w_j < \gamma\\
    0,              & \text{otherwise}
\end{cases}
\label{eqn:expert-dropout}
\end{equation}

\begin{wrapfigure}{r}{0.7\textwidth}
    \vspace{-17pt}
    \centering
    \includegraphics[width=\linewidth]{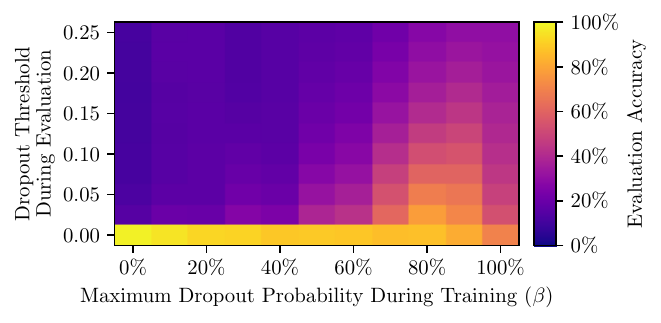}
    \caption{\textbf{Model accuracy after removing low-weighted experts across different training dropout levels.} If models trained without dropout ($\beta=0$) are prevented from using experts that contribute very little to the output, accuracy drops precipitously, from 98.2\% to 15.1\%. By comparison, the accuracy of models trained with a maximum dropout value of $\beta=0.8$ only drops from 86.5\% to 77.7\%.}
    \label{fig:expert-dropout}
    \vspace{-5pt}
\end{wrapfigure}

We set $\gamma$ to 0.1, meaning that experts contributing 10\% or more to the output of a layer are never deactivated. As this contribution weight decreases to zero, this probability increases linearly to $\beta$.
To identify how much dropout is needed to improve robustness, we train 11 groups of models with $\beta$ values ranging from $0, 0.1, ..., 0.9, 1$ using \cref{eqn:expert-dropout}, with 10 models in each group. We evaluate each model on 50 trials per task, blocking experts with routing weights below 11 values: $0, 0.025, \dots, 0.225, 0.25$. Accuracy is averaged across the 10 models in each group. We find that expert dropout has a relatively minor impact on performance, while significantly improving the robustness of the pathways that form. \cref{fig:expert-dropout} shows that for models trained with a maximum dropout of 80\% ($\beta=0.8$), this drop in accuracy is small (from 85.8\% to 74.4\%). This motivates us to set $\beta=0.8$ for our models in the remainder of this work. Note that routing consistency (\cref{sec:routing-stability}) remains high with dropout, shown through an average pairwise correlation of 0.51, which is significant over the baseline with $p<0.0001$ (see \cref{app:penalty_calc}).

\subsection{Distinct Pathways Across and Within Tasks}

For our final criterion, we want to identify whether meaningful patterns of expert usage develop across tasks and timescales within our model. To do this, we record the routing patterns across 50 trials of each task for both the baseline model and our final model, which is trained using the routing cost with task performance scaling and expert dropout. To visualize how routing varies across layers, tasks, and time, we average routing patterns for each task in three phases: (i) before the stimulus is shown, (ii) while the stimulus is shown, and (iii) during the response period. We apply K-means clustering ($k=10$) to these matrices to identify groups of tasks that use similar pathways.

\begin{figure}[ht]
    \centering
    \includegraphics[width=\linewidth]{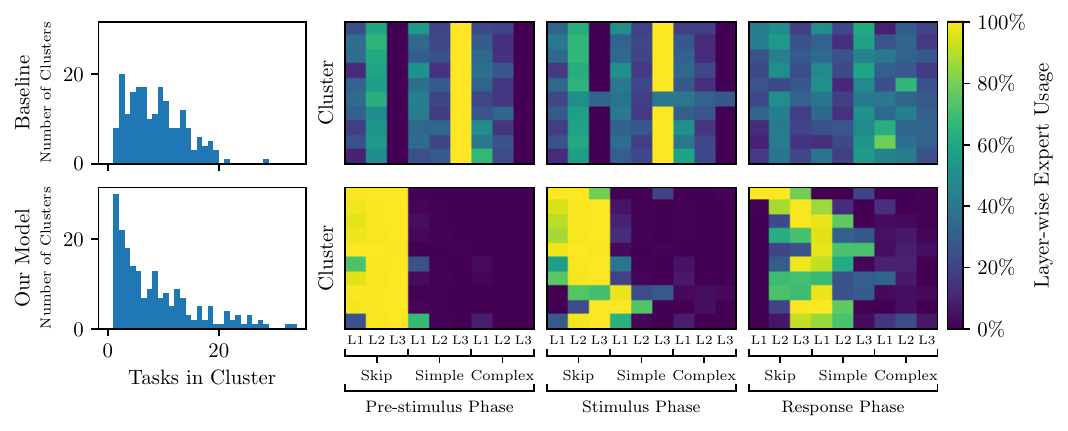}
    \caption{\textbf{Task clustering derived from expert usage patterns.} Clusters are averaged over three phases of each task: the pre-stimulus phase, during which the task is known but no input data is presented, the stimulus phase, during which input data is presented, and the response phase. Left: Sizes of clusters across training runs. Right: Average routing weights across training runs by cluster (y-axis) and task phase. In each phase, expert usages within each layer (across each layer's skip connection, simple expert, and complex expert) sum to 100\%. The complex expert is rarely used in the first two phases, during which the model outputs zero, but has an average usage as high as 11\% in the most complex cluster (see \cref{app:expert-usage}). L1 / L2 / L3 = Layer 1 / Layer 2 / Layer 3.}
    \label{fig:expert-usages-clustered}
\end{figure}

In our model, we observe a structured usage of expert pathways (\cref{fig:expert-usages-clustered}): during the pre-stimulus phase, models primarily rely on the cheap skip connections, as no information needs to be processed yet.
For some tasks, increasingly complex experts are activated with the onset of stimuli.
However, since the model still only needs to output zero during this phase, most tasks continue to leverage the cheap `all-skip' pathway until the response phase.
During this final phase, we see very rich dynamics of pathways being differentially activated across tasks and time periods in our model. This shows how processing pathways interact over the time of a trial, with different combinations of experts activated over tasks and time periods.
The following sections analyze these dynamics in more detail, especially in comparison to the dynamics observed in the brain.
Importantly, for the baseline model, clusters do not seem to employ very different combinations of experts across tasks.
The more differentiated usage of pathways across clusters can be seen in the distribution of the numbers of tasks contained in a given cluster (\cref{fig:expert-usages-clustered}, left).
Here, our model shows a very distinct power-law distribution of several large clusters containing $>20$ tasks and many small clusters capturing task-specific pathway usages.
The baseline model, on the other hand, seems to distribute the tasks more evenly across clusters, indicating that there are much less distinct routing patterns.
This can be shown quantitatively in the sizes of the largest clusters for each model run, which are significantly larger for our model than for the baseline model across 10 different random seeds ($p<0.0001$).
Further visualizations are provided in \cref{app:task-specific-expert-usage,app:expert-usage}.

Our results so far show that pathways do not automatically develop from a heterogeneous mixture of experts. Training models with a routing-complexity loss, scaling it based on task performance, and adding expert dropout, all encourage stable pathways to form. These three features define our \textit{Mixture-of-Pathways} (MoP) model. In the following section, we investigate whether the pathways that form in our MoP model mirror the pathways observed within the primate brain.
\section{Do Artificial Pathways Behave Like the Brain's Pathways?}
\label{sec:pathway-behavior}

In the previous section, we showed how our brain-inspired architectural contributions resulted in the formation of a mixture of processing pathways in our model.
Now, we will evaluate the degree to which these artificial pathways resemble the behavior of established processing pathways in the brain. Our analyses primarily focus on pathways and dynamics of the brain relating to task difficulty.

\subsection{Solving Tasks of Varying Difficulty}  
\label{sec:solving-tasks}

When analyzing brain activations across tasks with varying levels of difficulty, there is a distinct group of activations in a large frontoparietal network when solving complicated tasks. Since this network is activate while solving any difficult task, it was named the multiple-demand (MD) system~\cite{duncan2025construction, fedorenko2013broad}. It can be identified both in humans and non-human primates \cite{mitchell2016putative, mione2025neural}. With this in mind, we now want to test whether the selection of experts used to solve a task is indicative of task difficulty.

\begin{figure}
    \centering
    \begin{subfigure}[t]{0.25\textwidth}
        \centering
        \raisebox{0.3cm}{\includegraphics[width=\linewidth]{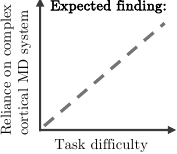}}
        \caption{Expected results}
    \end{subfigure}
    \hfill
    \begin{subfigure}[t]{0.74\textwidth}
        \centering
        \includegraphics[width=\linewidth]{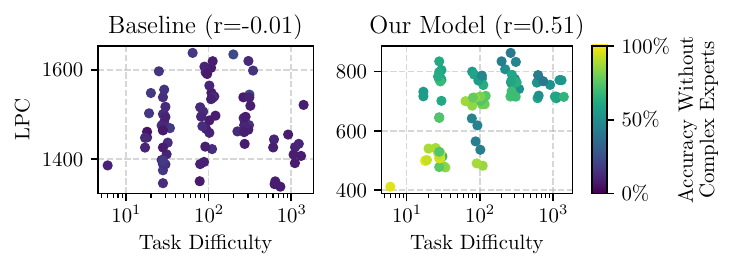}
        \caption{Empirical results}
    \end{subfigure}
    \caption{\textbf{Our model allocates less computation toward solving simpler tasks.} Additionally, when the most complex experts in each layer are disabled, our model is still able to solve the simplest tasks with high accuracy.}
    \label{fig:task-complexities}
\end{figure}

To relate these findings from the MD system to our model, we expect that when solving tasks of increasing difficulty, our model should learn to activate increasingly complex regions (schematic in \cref{fig:task-complexities}). We test this by measuring the correlation between a task's difficulty and the learned pathway complexity ($LPC$) used by the model to solve the task. We quantify the difficulty of a task by the number of training steps it takes a standalone GRU to learn the task (see \cref{app:task-difficulty} for details and alternative ways of quantifying task difficulty). We find that our full MoP model shows a positive significant relationship, whereas the baseline model does not, matching our expectations.

Furthermore, it is known that patients with lesions to their MD system struggle solving difficult tasks but their ability to solve simple tasks usually is unaffected \cite{roca_executive_2010,goel_are_1995}. We find that the same is true in our models: if the most complex expert in each layer is lesioned, our model can still solve simple tasks with high accuracy, but accuracy on difficulty tasks drops significantly. The same is not true of the baseline model: performance on all tasks drops to near-chance.

\subsection{Learning Tasks of Varying Difficulty}
\label{sec:learning-tasks}

\begin{figure}
    \vspace{-2mm}
    \centering
    \valign{#\cr
        \hsize=0.315\textwidth
        \begin{subfigure}{0.315\columnwidth}
            \centering
            \vspace{0.4cm}
            \includegraphics[width=1.1\linewidth]{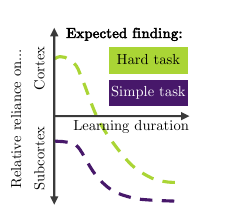}
            \vspace{0.9cm}
            \caption{Expected results}
            \label{fig:learning-phases}
        \end{subfigure}\cr\noalign{\hfill}
        \hsize=0.67\textwidth
        \begin{subfigure}{0.67\columnwidth}
            \centering
            \includegraphics[width=\linewidth]{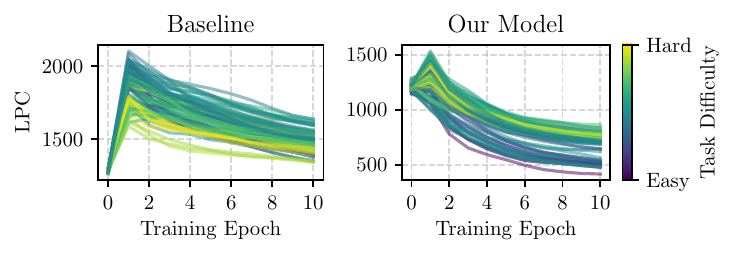}
        \end{subfigure}\vspace{-0.1cm}
        \begin{subfigure}{0.67\columnwidth}
            \hspace{0.13cm}\includegraphics[width=0.83\linewidth]{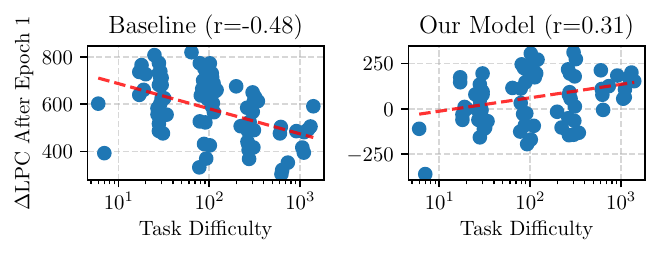}
            \caption{Empirical results}
            \label{fig:task-complexities-by-epoch-2}
        \end{subfigure}\cr
    }
    \caption{\textbf{Our model moves complex tasks to more complex pathways at the start of training to support the learning process.} (a) Conceptual schematic of cortical-subcortical interactions. (b) Top row shows the pathway complexity over learning per task averaged across 10 training runs. Lower row shows the change in pathway complexity between model initialization and the end of the first epoch as a function of task difficulty. Our model specifically relies on complex pathways to learn difficult tasks, similar to how the brain relies on complex pathways to support the acquisition of complex skills, even if they are gradually moved to simpler pathways later on.}
    \label{fig:lpc-over-time}
\end{figure}

A more nuanced view of pathways in the brain comes from observing how tasks of varying difficulty are learned over time. Here, an interesting distinction between complex and simple tasks is observed: while simple tasks can be learned through simpler (subcortical) regions alone, complex tasks require more complex (cortical) regions for learning~\cite{hong2018sensation,peters2021striatal,wolff_distinct_2022, dolan2013goals} (see schematic in \cref{fig:learning-phases}). However, as learning continues, even complex task skills are often ``transferred'' from complex to simple brain regions. 
This is possible despite the simple pathway not being sufficient to drive the learning process in the first place. 
We now want to test whether this phenomenon can be observed in our model.

To study this effect, we track the complexities of the learned pathways across tasks over the duration of learning. \cref{fig:lpc-over-time} shows these results across models: our MoP model seems to indeed learn complex tasks by first increasing their pathway complexity, but then reducing it gradually during learning. In contrast, very simple tasks do not increase in pathway complexity at all over learning. We now want to quantify this effect. To translate \cref{fig:learning-phases} to our models, we can quantify to which degree the pathway complexity of a given task is increasing or decreasing after the first training epoch. This is a measure of how much the model specifically uses a more complex pathway to learn a given task. Based on findings from neuroscience, we would expect complex tasks to be explicitly moved to more complex pathways, relative to the random starting point, whereas this should not be necessary for simple tasks \cite{kawai2015motor, hong2018sensation, mizes2024role}. \cref{fig:task-complexities-by-epoch-2} shows that this phenomenon can be observed in our model. The more difficult a task is, the more its pathway complexity increases at the start of learning, with the simplest tasks immediately getting routed toward simpler pathways ($r=0.31$, $p=0.0040$). This is not true in the baseline model, where we observe the opposite effect: the pathway complexity used to solve the most difficult tasks increases the least at the start of learning ($r=-0.48$, $p<0.0001$). Upon further inspection, this happens because in an effort to minimize the routing loss, the baseline model learns to push tasks that it is unable to solve toward simpler pathways prematurely. As a result, the baseline model fails to learn the most difficult tasks until the very end of the training process.

\subsection{Ablations}

Lastly, we investigate how changes made to our model can alter the degree to which it resembles pathways in the brain, as discussed in \cref{sec:solving-tasks,sec:learning-tasks}.
\cref{tab:ablations} shows the effects of design parameters on the correlations reported in \cref{fig:task-complexities,fig:lpc-over-time}. Notably, we found that when trained with our loss function that scales based on LPC but without dropout, our model exhibits the effect shown in \cref{fig:task-complexities}, but not the effect shown in \cref{fig:lpc-over-time}. This indicates that the finding in \cref{fig:lpc-over-time} is due to an interaction between the LPC scaling in our loss function and dropout, and can not be explained by training with the LPC regularization alone. We speculate that expert dropout forces the model to be explicit about which pathway is used in learning, and is crucial for creating brain-like learning dynamics.
This finding highlights how our model's behavior specifically results from the interplay of all three of our proposed inductive biases.

We also find that removing the task embedding layer improves the finding in \cref{fig:lpc-over-time}, however its removal drastically slows down training since the active task is represented with 82 dimensions at each timestep rather than 16 (see \cref{sec:methods-training}).
Changing the router's hidden size does not meaningfully affect our results.
Increasing or decreasing the penalty for using large experts ($\alpha$) naturally has a meaningful effect on the results, where too strong of a penalty inhibits learning the tasks as well as general convergence of the model, and too weak of a penalty does not sufficiently motivate the model to reduce its usage of complex experts.
All rows in the table are averaged over 10 runs.

\begin{table}[t]
\caption{Task accuracy and pathway metrics for modified versions of our model.}
\vspace{0.1cm}
\centering
\begin{threeparttable}
\begin{tabular}{lrrr}
\toprule
\textbf{Model} & \textbf{Accuracy} & \textbf{Fig. 5 Correlation} & \textbf{Fig. 6b Correlation} \\
\midrule
Baseline & 91.1\% $\pm$ 8.9\% & -0.01 & -0.49$^{***}$ \\
Our model$^{\dagger}$ & 83.0\% $\pm$ 15.5\% & 0.54$^{***}$ & 0.31$^{**}$ \\
\midrule
Without dropout & 90.1\% $\pm$ 8.9\% & 0.55$^{***}$ & 0.03 \\
$\alpha=1e^{-4}$ & 69.0\% $\pm$ 20.1\% & -0.57$^{***}$ & -0.37$^{***}$ \\
$\alpha=1e^{-6}$ & 89.7\% $\pm$ 9.0\% & \textbf{0.62$^{***}$} & 0.18 \\
Without task embeddings & 83.0\% $\pm$ 14.2\% & 0.58$^{***}$ & \textbf{0.58$^{***}$} \\
Router dim = 32 & 83.2\% $\pm$ 14.5\% & 0.46$^{***}$ & 0.33$^{**}$ \\
Router dim = 128 & 81.9\% $\pm$ 16.9\% & 0.35$^{**}$ & 0.27$^{*}$ \\
\bottomrule
\end{tabular}
\begin{tablenotes}
\footnotesize
\item $\dagger$ With dropout, $\alpha=1e^{-5}$, task embeddings, and router dim = 64
\item * $p<0.05$, ** $p<0.01$, *** $p<0.001$
\end{tablenotes}
\end{threeparttable}
\label{tab:ablations}
\end{table}
\section{Discussion}
\label{sec:discussion}

In this work, we adapted the heterogeneous Mixture-of-Experts architecture to investigate how brain-like processing pathways can form between layers of heterogeneous experts. While these experts do not form pathways on their own, once trained with a routing-cost loss, task-performance scaling, and expert dropout, we find that they create a \textit{Mixture of Pathways}. Our model provides an account of task-specific brain-wide pathways commonly observed in neuroscience.

\textbf{These findings are relevant for neuroscience}, as energetic and processing complexity related priors have been key explanatory mechanisms for how the structure and function of the brain arises \cite{bullmore2012economy, achterberg_spatially_2023, akarca2023weighted, kool_mental_2018}. We show that incentivizing models to learn to prioritize simple experts drives the development of brain-like processing pathways from a heterogeneous set of expert models. Additionally, we show how stochasticity of signals is important for learning self-sufficient processing networks. Our model represents an exciting new architecture which can be expanded in the future to study additional heterogeneities present in the connectome, such as varying cell-types and neurotransmitters. Region-specific models, such as those for the hippocampus~\cite{chandra2025episodic}, could be integrated within our architecture. 

The brain's implementation of the complexity-guided routing mechanism would likely be found in thalamic nuclei, which regulate information flow between cortical regions \cite{sherman_role_2002}. Two systems could modulate pathway selection based on metabolic costs: norepinephrine release from the locus coeruleus, which controls cognitive effort and processing power allocation \cite{aston2005integrative, sara2012orienting, westbrook2015cognitive}, and hypocretin/orexin neurons in the hypothalamus, which govern metabolic resource budgets \cite{tesmer2025neurometabolic}. Both systems project strongly to thalamic nuclei and could influence routing decisions between simple and complex processing pathways. This suggests our model's routing-cost mechanism may reflect how the brain balances computational demands against metabolic constraints through neuromodulatory control of thalamocortical interactions. While mapping our router onto a specific brain region might feel natural, it should be added that mechanisms like predictive coding can implement routing and filtering operation between regions without the need of an explicit router region \cite{gabhart2023predictive}. 

\textbf{In the context of machine learning}, our work builds on the recent widespread adoption of the Mixture-of-Experts architecture for building parameter-efficient large language models \cite{liu2024deepseek, llama_team_llama_2025}. Recent innovations specifically aim at allowing MoE models to process queries dynamically to reduce processing costs \cite{raposo2024mixture}. Our small-scale simulations show how it may be possible to use heterogeneous experts alongside a processing-cost loss function that allows the model to dynamically allocate resources to processing tokens. Finally, load balancing in MoEs prevents over-reliance on a single expert by encouraging distributed processing \cite{liu2024deepseek}. Our complexity loss serves as a task-driven form of load balancing.

\subsection{Limitations and Extensions}
\label{sec:architecture_extensions}
\label{sec:limitations}

There are several ways to expand our investigations. \textbf{On the neuroscience side}, our architecture introduces a new way of modeling multi-region interactions of the brain, but some key architectural characteristics are not yet taken into account. Most importantly, the primate brain has large loop structures which would allow signals to return to a region \cite{achterberg2023building}. Our architecture only allows a forward progression of signal and does not allow signals to be routed back to earlier layers. At the same time, while our analyses demonstrated a link between experts and cortical and subcortical regions, we have not linked the router component of our HMoE layers to a specific component of the brain. Potential options are discussed earlier in \cref{sec:discussion}, we do not make any explicit comparison to brain data yet. \textbf{On the ML side}, our training setup currently focuses on solving relatively simple tasks with a small model. To see whether our complexity-aware routing and load-balancing measures scale to larger networks, we would need to train larger models on more difficult tasks. Lastly, \textbf{on the identification of pathways}, we currently rely on three independent tests to see whether a model contains pathways. Future investigations would ideally identify one specific metric to quantify the degree to which a Mixture-of-Experts architecture has formed pathways.
\section{Conclusion}

In this paper, we introduced a modified Heterogeneous Mixture-of-Experts architecture that results in the formation of recognizable processing pathways. Analysis of these pathways during learning and problem solving revealed a similarity between these pathways and those observed in the brain. Our new \textit{Mixture-of-Pathways} architecture serves as a new theoretical tool for neuroscience and can guide the search for future resource-efficient architectures in machine learning.

\newpage

\begin{ack}
We thank our reviewers, the \textit{Neural \& Machine Learning Group} at the University of Oxford, and the \textit{AI HW SW CoDesign Workstream} at the Open Compute Project (OCP) for helpful feedback.
This research is supported by the EPSRC (EP/X029336/1) and an ERC-UKRA Frontier Research Guarantee Starting Grant (EP/Y027841/1) awarded to R.P.C.
J.C.'s work was supported by a Rhodes Scholarship.
J.A.'s work was partially supported by a Career Development Research Fellowship from St John's College, Oxford.
We additionally thank Modal for compute credits.
\end{ack}

\bibliographystyle{unsrt}
\bibliography{ref,ref_revision}

@article{achterberg2023building,
  title={Building artificial neural circuits for domain-general cognition: a primer on brain-inspired systems-level architecture},
  author={Achterberg, Jascha and Akarca, Danyal and Assem, Moataz and Heimbach, Moritz and Astle, Duncan E and Duncan, John},
  journal={arXiv preprint arXiv:2303.13651},
  year={2023}
}

@article{defazio2024road,
  title={The road less scheduled},
  author={Defazio, Aaron and Yang, Xingyu and Khaled, Ahmed and Mishchenko, Konstantin and Mehta, Harsh and Cutkosky, Ashok},
  journal={Advances in Neural Information Processing Systems},
  volume={37},
  pages={9974--10007},
  year={2024}
}

@article{wang2024hmoe,
  title={Hmoe: Heterogeneous mixture of experts for language modeling},
  author={Wang, An and Sun, Xingwu and Xie, Ruobing and Li, Shuaipeng and Zhu, Jiaqi and Yang, Zhen and Zhao, Pinxue and Han, JN and Kang, Zhanhui and Wang, Di and others},
  journal={arXiv preprint arXiv:2408.10681},
  year={2024}
}

@article{raposo2024mixture,
  title={Mixture-of-depths: Dynamically allocating compute in transformer-based language models},
  author={Raposo, David and Ritter, Sam and Richards, Blake and Lillicrap, Timothy and Humphreys, Peter Conway and Santoro, Adam},
  journal={arXiv preprint arXiv:2404.02258},
  year={2024}
}

@article{gabhart2023predictive,
  title={Predictive coding: a more cognitive process than we thought?},
  author={Gabhart, Kaitlyn M and Xiong, Yihan Sophy and Bastos, Andr{\'e} M},
  journal={Trends in Cognitive Sciences},
  year={2023},
  publisher={Elsevier}
}

@article{olson2025semantic,
  title={Semantic Specialization in MoE Appears with Scale: A Study of DeepSeek R1 Expert Specialization},
  author={Olson, Matthew Lyle and Ratzlaff, Neale and Hinck, Musashi and Luo, Man and Yu, Sungduk and Xue, Chendi and Lal, Vasudev},
  journal={arXiv preprint arXiv:2502.10928},
  year={2025}
}

@article{kozachkov2022rnns,
  title={RNNs of RNNs: Recursive construction of stable assemblies of recurrent neural networks},
  author={Kozachkov, Leo and Ennis, Michaela and Slotine, Jean-Jacques},
  journal={Advances in neural information processing systems},
  volume={35},
  pages={30512--30527},
  year={2022}
}

@article{pereira2024cognitive,
  title={Cognitive network interactions through communication subspaces in large-scale models of the neocortex},
  author={Pereira-Obilinovic, Ulises and Froudist-Walsh, Sean and Wang, Xiao-Jing},
  journal={bioRxiv},
  year={2024}
}

@article{clark2025structure,
  title={Structure of activity in multiregion recurrent neural networks},
  author={Clark, David G and Beiran, Manuel},
  journal={Proceedings of the National Academy of Sciences},
  volume={122},
  number={10},
  pages={e2404039122},
  year={2025},
  publisher={National Academy of Sciences}
}

@article{yang2021towards,
  title={Towards the next generation of recurrent network models for cognitive neuroscience},
  author={Yang, Guangyu Robert and Molano-Maz{\'o}n, Manuel},
  journal={Current opinion in neurobiology},
  volume={70},
  pages={182--192},
  year={2021},
  publisher={Elsevier}
}

@article{mizes2024role,
  title={The role of motor cortex in motor sequence execution depends on demands for flexibility},
  author={Mizes, Kevin GC and Lindsey, Jack and Escola, G Sean and {\"O}lveczky, Bence P},
  journal={Nature Neuroscience},
  pages={1--10},
  year={2024},
  publisher={Nature Publishing Group US New York}
}

@article{zhou2021distributed,
  title={Distributed functions of prefrontal and parietal cortices during sequential categorical decisions},
  author={Zhou, Yang and Rosen, Matthew C and Swaminathan, Sruthi K and Masse, Nicolas Y and Zhu, Ou and Freedman, David J},
  journal={Elife},
  volume={10},
  pages={e58782},
  year={2021},
  publisher={eLife Sciences Publications Limited}
}

@article{finzi2023single,
  title={A single computational objective drives specialization of streams in visual cortex},
  author={Finzi, Dawn and Margalit, Eshed and Kay, Kendrick and Yamins, Daniel LK and Grill-Spector, Kalanit},
  journal={bioRxiv},
  pages={2023--12},
  year={2023},
  publisher={Cold Spring Harbor Laboratory}
}

@article{pemberton2024cerebellar,
  title={Cerebellar-driven cortical dynamics can enable task acquisition, switching and consolidation},
  author={Pemberton, Joseph and Chadderton, Paul and Costa, Rui Ponte},
  journal={Nature Communications},
  volume={15},
  number={1},
  pages={10913},
  year={2024},
  publisher={Nature Publishing Group UK London}
}

@article{fang2023predictive,
  title={Predictive auxiliary objectives in deep rl mimic learning in the brain},
  author={Fang, Ching and Stachenfeld, Kimberly L},
  journal={arXiv preprint arXiv:2310.06089},
  year={2023}
}

@article{moskovitz2024understanding,
  title={Understanding dual process cognition via the minimum description length principle},
  author={Moskovitz, Ted and Miller, Kevin J and Sahani, Maneesh and Botvinick, Matthew M},
  journal={PLOS Computational Biology},
  volume={20},
  number={10},
  pages={e1012383},
  year={2024},
  publisher={Public Library of Science San Francisco, CA USA}
}

@article{liebana2023striatal,
  title={Striatal dopamine reflects individual long-term learning trajectories},
  author={Liebana Garcia, Samuel and Laffere, Aeron and Toschi, Chiara and Schilling, Louisa and Podlaski, Jacek and Fritsche, Matthias and Zatka-Haas, Peter and Li, Yulong and Bogacz, Rafal and Saxe, Andrew and others},
  journal={bioRxiv},
  pages={2023--12},
  year={2023},
  publisher={Cold Spring Harbor Laboratory}
}

@article{liu2024deepseek,
  title={Deepseek-v2: A strong, economical, and efficient mixture-of-experts language model},
  author={Liu, Aixin and Feng, Bei and Wang, Bin and Wang, Bingxuan and Liu, Bo and Zhao, Chenggang and Dengr, Chengqi and Ruan, Chong and Dai, Damai and Guo, Daya and others},
  journal={arXiv preprint arXiv:2405.04434},
  year={2024}
}

@article{jawahar2022automoe,
  title={Automoe: Heterogeneous mixture-of-experts with adaptive computation for efficient neural machine translation},
  author={Jawahar, Ganesh and Mukherjee, Subhabrata and Liu, Xiaodong and Kim, Young Jin and Abdul-Mageed, Muhammad and Lakshmanan, Laks VS and Awadallah, Ahmed Hassan and Bubeck, Sebastien and Gao, Jianfeng},
  journal={arXiv preprint arXiv:2210.07535},
  year={2022}
}

@article{yang2024vision,
  title={A vision chip with complementary pathways for open-world sensing},
  author={Yang, Zheyu and Wang, Taoyi and Lin, Yihan and Chen, Yuguo and Zeng, Hui and Pei, Jing and Wang, Jiazheng and Liu, Xue and Zhou, Yichun and Zhang, Jianqiang and others},
  journal={Nature},
  volume={629},
  number={8014},
  pages={1027--1033},
  year={2024},
  publisher={Nature Publishing Group UK London}
}

@article{duncan2025construction,
  title={Construction and use of mental models: Organizing principles for the science of brain and mind},
  author={Duncan, John},
  journal={Neuropsychologia},
  volume={207},
  pages={109062},
  year={2025},
  publisher={Elsevier}
}

@article{de2011usefulness,
  title={On the usefulness of ‘what’and ‘where’pathways in vision},
  author={de Haan, Edward HF and Cowey, Alan},
  journal={Trends in cognitive sciences},
  volume={15},
  number={10},
  pages={460--466},
  year={2011},
  publisher={Elsevier}
}

@article{arnott2004assessing,
  title={Assessing the auditory dual-pathway model in humans},
  author={Arnott, Stephen R and Binns, Malcolm A and Grady, Cheryl L and Alain, Claude},
  journal={Neuroimage},
  volume={22},
  number={1},
  pages={401--408},
  year={2004},
  publisher={Elsevier}
}

@article{bernstein2015two,
  title={Two neural pathways of face processing: A critical evaluation of current models},
  author={Bernstein, Michal and Yovel, Galit},
  journal={Neuroscience \& Biobehavioral Reviews},
  volume={55},
  pages={536--546},
  year={2015},
  publisher={Elsevier}
}

@article{dolan2013goals,
  title={Goals and habits in the brain},
  author={Dolan, Ray J and Dayan, Peter},
  journal={Neuron},
  volume={80},
  number={2},
  pages={312--325},
  year={2013},
  publisher={Elsevier}
}

@article{peters2021striatal,
  title={Striatal activity topographically reflects cortical activity},
  author={Peters, Andrew J and Fabre, Julie MJ and Steinmetz, Nicholas A and Harris, Kenneth D and Carandini, Matteo},
  journal={Nature},
  volume={591},
  number={7850},
  pages={420--425},
  year={2021},
  publisher={Nature Publishing Group UK London}
}

@article{hong2018sensation,
  title={Sensation, movement and learning in the absence of barrel cortex},
  author={Hong, Y Kate and Lacefield, Clay O and Rodgers, Chris C and Bruno, Randy M},
  journal={Nature},
  volume={561},
  number={7724},
  pages={542--546},
  year={2018},
  publisher={Nature Publishing Group UK London}
}

@article{yao2023high,
  title={A high-resolution transcriptomic and spatial atlas of cell types in the whole mouse brain},
  author={Yao, Zizhen and van Velthoven, Cindy TJ and Kunst, Michael and Zhang, Meng and McMillen, Delissa and Lee, Changkyu and Jung, Won and Goldy, Jeff and Abdelhak, Aliya and Aitken, Matthew and others},
  journal={Nature},
  volume={624},
  number={7991},
  pages={317--332},
  year={2023},
  publisher={Nature Publishing Group UK London}
}

@article{glasser2016multi,
  title={A multi-modal parcellation of human cerebral cortex},
  author={Glasser, Matthew F and Coalson, Timothy S and Robinson, Emma C and Hacker, Carl D and Harwell, John and Yacoub, Essa and Ugurbil, Kamil and Andersson, Jesper and Beckmann, Christian F and Jenkinson, Mark and others},
  journal={Nature},
  volume={536},
  number={7615},
  pages={171--178},
  year={2016},
  publisher={Nature Publishing Group}
}

@article{shipp2007structure,
  title={Structure and function of the cerebral cortex},
  author={Shipp, Stewart},
  journal={Current Biology},
  volume={17},
  number={12},
  pages={R443--R449},
  year={2007},
  publisher={Elsevier}
}

@article{bertolero2015modular,
  title={The modular and integrative functional architecture of the human brain},
  author={Bertolero, Maxwell A and Yeo, BT Thomas and D’Esposito, Mark},
  journal={Proceedings of the National Academy of Sciences},
  volume={112},
  number={49},
  pages={E6798--E6807},
  year={2015},
  publisher={National Academy of Sciences}
}

@article{yang2019task,
  title={Task representations in neural networks trained to perform many cognitive tasks},
  author={Yang, Guangyu Robert and Joglekar, Madhura R and Song, H Francis and Newsome, William T and Wang, Xiao-Jing},
  journal={Nature neuroscience},
  volume={22},
  number={2},
  pages={297--306},
  year={2019},
  publisher={Nature Publishing Group US New York}
}

@article{sporns2016modular,
  title={Modular brain networks},
  author={Sporns, Olaf and Betzel, Richard F},
  journal={Annual review of psychology},
  volume={67},
  number={1},
  pages={613--640},
  year={2016},
  publisher={Annual Reviews}
}

@article{fedus2022review,
  title={A review of sparse expert models in deep learning},
  author={Fedus, William and Dean, Jeff and Zoph, Barret},
  journal={arXiv preprint arXiv:2209.01667},
  year={2022}
}

@article{barham2022pathways,
  title={Pathways: Asynchronous distributed dataflow for ml},
  author={Barham, Paul and Chowdhery, Aakanksha and Dean, Jeff and Ghemawat, Sanjay and Hand, Steven and Hurt, Daniel and Isard, Michael and Lim, Hyeontaek and Pang, Ruoming and Roy, Sudip and others},
  journal={Proceedings of Machine Learning and Systems},
  volume={4},
  pages={430--449},
  year={2022}
}

@article{gritsch2024nexus,
  title={Nexus: Specialization meets Adaptability for Efficiently Training Mixture of Experts},
  author={Gritsch, Nikolas and Zhang, Qizhen and Locatelli, Acyr and Hooker, Sara and {\"U}st{\"u}n, Ahmet},
  journal={arXiv preprint arXiv:2408.15901},
  year={2024}
}

@article{he2024expertflow,
  title={Expertflow: Optimized expert activation and token allocation for efficient mixture-of-experts inference},
  author={He, Xin and Zhang, Shunkang and Wang, Yuxin and Yin, Haiyan and Zeng, Zihao and Shi, Shaohuai and Tang, Zhenheng and Chu, Xiaowen and Tsang, Ivor and Soon, Ong Yew},
  journal={arXiv preprint arXiv:2410.17954},
  year={2024}
}

@article{perez2021neural,
  title={Neural heterogeneity promotes robust learning},
  author={Perez-Nieves, Nicolas and Leung, Vincent CH and Dragotti, Pier Luigi and Goodman, Dan FM},
  journal={Nature communications},
  volume={12},
  number={1},
  pages={5791},
  year={2021},
  publisher={Nature Publishing Group UK London}
}

@article{bena2025dynamics,
  title={Dynamics of specialization in neural modules under resource constraints},
  author={B{\'e}na, Gabriel and Goodman, Dan FM},
  journal={Nature Communications},
  volume={16},
  number={1},
  pages={187},
  year={2025},
  publisher={Nature Publishing Group UK London}
}

@article{sheeran2024spatial,
  title={Spatial embedding promotes a specific form of modularity with low entropy and heterogeneous spectral dynamics},
  author={Sheeran, Cornelia and Ham, Andrew S and Astle, Duncan E and Achterberg, Jascha and Akarca, Danyal},
  journal={arXiv preprint arXiv:2409.17693},
  year={2024}
}

@article{stroud2025effects,
  title={Effects of noise and metabolic cost on cortical task representations},
  author={Stroud, Jake Patrick and Wojcik, Michal and Jensen, Kristopher Torp and Kusunoki, Makoto and Kadohisa, Mikiko and Buckley, Mark J and Duncan, John and Stokes, Mark G and Lengyel, M{\'a}t{\'e}},
  journal={eLife},
  volume={13},
  pages={RP94961},
  year={2025},
  publisher={eLife Sciences Publications Limited}
}

@article{akarca2023weighted,
  title={A weighted generative model of the human connectome},
  author={Akarca, Danyal and Schiavi, Simona and Achterberg, Jascha and Genc, Sila and Jones, Derek K and Astle, Duncan E},
  journal={bioRxiv},
  pages={2023--06},
  year={2023},
  publisher={Cold Spring Harbor Laboratory}
}

@article{gershman2015computational,
  title={Computational rationality: A converging paradigm for intelligence in brains, minds, and machines},
  author={Gershman, Samuel J and Horvitz, Eric J and Tenenbaum, Joshua B},
  journal={Science},
  volume={349},
  number={6245},
  pages={273--278},
  year={2015},
  publisher={American Association for the Advancement of Science}
}

@article{ali2022predictive,
  title={Predictive coding is a consequence of energy efficiency in recurrent neural networks},
  author={Ali, Abdullahi and Ahmad, Nasir and de Groot, Elgar and van Gerven, Marcel Antonius Johannes and Kietzmann, Tim Christian},
  journal={Patterns},
  volume={3},
  number={12},
  year={2022},
  publisher={Elsevier}
}

@article{molano2022neurogym,
  title={NeuroGym: An open resource for developing and sharing neuroscience tasks},
  author={Molano-Mazon, Manuel and Barbosa, Joao and Pastor-Ciurana, Jordi and Fradera, Marta and Zhang, Ru-Yuan and Forest, Jeremy and del Pozo Lerida, Jorge and Ji-An, Li and Cueva, Christopher J and de la Rocha, Jaime and others},
  year={2022},
  journal={OSF}
}

@article{bullmore2012economy,
  title={The economy of brain network organization},
  author={Bullmore, Ed and Sporns, Olaf},
  journal={Nature reviews neuroscience},
  volume={13},
  number={5},
  pages={336--349},
  year={2012},
  publisher={Nature Publishing Group UK London}
}

@article{aimone2021roadmap,
  title={A roadmap for reaching the potential of brain-derived computing},
  author={Aimone, James B},
  journal={Advanced Intelligent Systems},
  volume={3},
  number={1},
  pages={2000191},
  year={2021},
  publisher={Wiley Online Library}
}

@article{li2021adaptive,
  title={Adaptive dropout method based on biological principles},
  author={Li, Hailiang and Weng, Jian and Mao, Yijun and Wang, Yonghua and Zhan, Yiju and Cai, Qingling and Gu, Wanrong},
  journal={IEEE Transactions on Neural Networks and Learning Systems},
  volume={32},
  number={9},
  pages={4267--4276},
  year={2021},
  publisher={IEEE}
}

@article{fedorenko2013broad,
  title={Broad domain generality in focal regions of frontal and parietal cortex},
  author={Fedorenko, Evelina and Duncan, John and Kanwisher, Nancy},
  journal={Proceedings of the National Academy of Sciences},
  volume={110},
  number={41},
  pages={16616--16621},
  year={2013},
  publisher={National Academy of Sciences}
}

@article{mitchell2016putative,
  title={A putative multiple-demand system in the macaque brain},
  author={Mitchell, Daniel J and Bell, Andrew H and Buckley, Mark J and Mitchell, Anna S and Sallet, Jerome and Duncan, John},
  journal={Journal of Neuroscience},
  volume={36},
  number={33},
  pages={8574--8585},
  year={2016},
  publisher={Society for Neuroscience}
}

@article{mione2025neural,
  title={Neural dynamics of an extended frontal lobe network in goal-subgoal problem solving},
  author={Mione, Valentina and Achterberg, Jascha and Kusunoki, Makoto and Buckley, Mark J and Duncan, John},
  journal={bioRxiv},
  pages={2025--05},
  year={2025},
  publisher={Cold Spring Harbor Laboratory}
}

@article{chandra2025episodic,
  title={Episodic and associative memory from spatial scaffolds in the hippocampus},
  author={Chandra, Sarthak and Sharma, Sugandha and Chaudhuri, Rishidev and Fiete, Ila},
  journal={Nature},
  pages={1--13},
  year={2025},
  publisher={Nature Publishing Group UK London}
}

@article{kawai2015motor,
  title={Motor cortex is required for learning but not for executing a motor skill},
  author={Kawai, Risa and Markman, Timothy and Poddar, Rajesh and Ko, Raymond and Fantana, Antoniu L and Dhawale, Ashesh K and Kampff, Adam R and {\"O}lveczky, Bence P},
  journal={Neuron},
  volume={86},
  number={3},
  pages={800--812},
  year={2015},
  publisher={Elsevier}
}

@article{sherman_role_2002,
	title = {The role of the thalamus in the flow of information to the cortex},
	volume = {357},
	url = {https://royalsocietypublishing.org/doi/abs/10.1098/rstb.2002.1161},
	doi = {10.1098/rstb.2002.1161},
	number = {1428},
	urldate = {2025-04-11},
	journal = {Philosophical Transactions of the Royal Society of London. Series B: Biological Sciences},
	author = {Sherman, S. M. and Guillery, R. W.},
	year = {2002},
	note = {Publisher: Royal Society},
	pages = {1695--1708},
}

@article{goel_are_1995,
	title = {Are the frontal lobes implicated in "planning" functions? {Interpreting} data from the {Tower} of {Hanoi}},
	volume = {33},
	issn = {0028-3932},
	shorttitle = {Are the frontal lobes implicated in "planning" functions?},
	doi = {10.1016/0028-3932(95)90866-p},
	language = {eng},
	number = {5},
	journal = {Neuropsychologia},
	author = {Goel, V. and Grafman, J.},
	year = {1995},
	pmid = {7637857},
	pages = {623--642},
}

@article{roca_executive_2010,
	title = {Executive function and fluid intelligence after frontal lobe lesions},
	volume = {133},
	issn = {0006-8950},
	url = {https://doi.org/10.1093/brain/awp269},
	doi = {10.1093/brain/awp269},
	number = {1},
	urldate = {2025-04-11},
	journal = {Brain},
	author = {Roca, María and Parr, Alice and Thompson, Russell and Woolgar, Alexandra and Torralva, Teresa and Antoun, Nagui and Manes, Facundo and Duncan, John},
	year = {2010},
	pages = {234--247},
}

@article{kool_mental_2018,
	title = {Mental labour},
	volume = {2},
	copyright = {2018 The Author(s)},
	issn = {2397-3374},
	url = {https://www.nature.com/articles/s41562-018-0401-9},
	doi = {10.1038/s41562-018-0401-9},
	language = {en},
	number = {12},
	urldate = {2025-04-11},
	journal = {Nature Human Behaviour},
	author = {Kool, Wouter and Botvinick, Matthew},
	year = {2018},
	note = {Publisher: Nature Publishing Group},
	pages = {899--908},
}

@article{achterberg_spatially_2023,
	title = {Spatially embedded recurrent neural networks reveal widespread links between structural and functional neuroscience findings},
	volume = {5},
	copyright = {2023 The Author(s)},
	issn = {2522-5839},
	url = {https://www.nature.com/articles/s42256-023-00748-9},
	doi = {10.1038/s42256-023-00748-9},
	language = {en},
	number = {12},
	urldate = {2025-04-11},
	journal = {Nature Machine Intelligence},
	author = {Achterberg, Jascha and Akarca, Danyal and Strouse, D. J. and Duncan, John and Astle, Duncan E.},
	year = {2023},
	note = {Publisher: Nature Publishing Group},
	pages = {1369--1381},
}

@article{deco_stochastic_2009,
	title = {Stochastic dynamics as a principle of brain function},
	volume = {88},
	issn = {0301-0082},
	url = {https://www.sciencedirect.com/science/article/pii/S0301008209000197},
	doi = {10.1016/j.pneurobio.2009.01.006},
	number = {1},
	urldate = {2025-04-11},
	journal = {Progress in Neurobiology},
	author = {Deco, Gustavo and Rolls, Edmund T. and Romo, Ranulfo},
	year = {2009},
	pages = {1--16},
}

@misc{llama_team_llama_2025,
	title = {The {Llama} 4 herd: {The} beginning of a new era of natively multimodal {AI} innovation},
	shorttitle = {The {Llama} 4 herd},
	url = {https://ai.meta.com/blog/llama-4-multimodal-intelligence/},
	language = {en},
	urldate = {2025-04-10},
	journal = {Meta AI},
	author = {{Llama Team}},
	year = {2025},
}

@article{etkin_emotional_2011,
	title = {Emotional processing in anterior cingulate and medial prefrontal cortex},
	volume = {15},
	issn = {1879-307X},
	doi = {10.1016/j.tics.2010.11.004},
	language = {eng},
	number = {2},
	journal = {Trends in Cognitive Sciences},
	author = {Etkin, Amit and Egner, Tobias and Kalisch, Raffael},
	year = {2011},
	pmid = {21167765},
	pmcid = {PMC3035157},
	pages = {85--93},
}

@article{grill-spector_human_2004,
	title = {The human visual cortex},
	volume = {27},
	issn = {0147-006X},
	doi = {10.1146/annurev.neuro.27.070203.144220},
	language = {eng},
	journal = {Annual Review of Neuroscience},
	author = {Grill-Spector, Kalanit and Malach, Rafael},
	year = {2004},
	pmid = {15217346},
	pages = {649--677},
}

@article{duncan_multiple-demand_2010,
	title = {The multiple-demand ({MD}) system of the primate brain: mental programs for intelligent behaviour},
	volume = {14},
	issn = {1364-6613},
	shorttitle = {The multiple-demand ({MD}) system of the primate brain},
	url = {https://www.sciencedirect.com/science/article/pii/S1364661310000057},
	doi = {10.1016/j.tics.2010.01.004},
	number = {4},
	urldate = {2025-04-09},
	journal = {Trends in Cognitive Sciences},
	author = {Duncan, John},
	year = {2010},
	pages = {172--179},
}

@article{wolff_distinct_2022,
	title = {Distinct roles for motor cortical and thalamic inputs to striatum during motor skill learning and execution},
	volume = {8},
	issn = {2375-2548},
	url = {https://www.ncbi.nlm.nih.gov/pmc/articles/PMC8880788/},
	doi = {10.1126/sciadv.abk0231},
	number = {8},
	urldate = {2025-04-09},
	journal = {Science Advances},
	author = {Wolff, Steffen B. E. and Ko, Raymond and Ölveczky, Bence P.},
	year = {2022},
	pmid = {35213216},
	pmcid = {PMC8880788},
	pages = {eabk0231},
}

@article{srivastava_dropout_2014,
	title = {Dropout: {A} {Simple} {Way} to {Prevent} {Neural} {Networks} from {Overfitting}},
	volume = {15},
	issn = {1533-7928},
	shorttitle = {Dropout},
	url = {http://jmlr.org/papers/v15/srivastava14a.html},
	number = {56},
	urldate = {2025-04-09},
	journal = {Journal of Machine Learning Research},
	author = {Srivastava, Nitish and Hinton, Geoffrey and Krizhevsky, Alex and Sutskever, Ilya and Salakhutdinov, Ruslan},
	year = {2014},
	pages = {1929--1958},
}

@misc{khona_winning_2023,
	title = {Winning the lottery with neural connectivity constraints: faster learning across cognitive tasks with spatially constrained sparse {RNNs}},
	shorttitle = {Winning the lottery with neural connectivity constraints},
	url = {http://arxiv.org/abs/2207.03523},
	doi = {10.48550/arXiv.2207.03523},
	urldate = {2025-04-08},
	publisher = {arXiv},
	author = {Khona, Mikail and Chandra, Sarthak and Ma, Joy J. and Fiete, Ila},
	year = {2023},
	note = {arXiv:2207.03523 [q-bio]},
}

@article{duncan_construction_2025,
	title = {Construction and use of mental models: {Organizing} principles for the science of brain and mind},
	volume = {207},
	issn = {0028-3932},
	shorttitle = {Construction and use of mental models},
	url = {https://www.sciencedirect.com/science/article/pii/S002839322400277X},
	doi = {10.1016/j.neuropsychologia.2024.109062},
	urldate = {2025-04-07},
	journal = {Neuropsychologia},
	author = {Duncan, John},
	year = {2025},
	pages = {109062},
}

@misc{fedus_switch_2022,
	title = {Switch {Transformers}: {Scaling} to {Trillion} {Parameter} {Models} with {Simple} and {Efficient} {Sparsity}},
	shorttitle = {Switch {Transformers}},
	url = {http://arxiv.org/abs/2101.03961},
	doi = {10.48550/arXiv.2101.03961},
	urldate = {2025-04-06},
	publisher = {arXiv},
	author = {Fedus, William and Zoph, Barret and Shazeer, Noam},
	year = {2022},
	note = {arXiv:2101.03961 [cs]},
}

@misc{jiang_mixtral_2024,
	title = {Mixtral of {Experts}},
	url = {http://arxiv.org/abs/2401.04088},
	doi = {10.48550/arXiv.2401.04088},
	urldate = {2025-04-06},
	publisher = {arXiv},
	author = {Jiang, Albert Q. and Sablayrolles, Alexandre and Roux, Antoine and Mensch, Arthur and Savary, Blanche and Bamford, Chris and Chaplot, Devendra Singh and Casas, Diego de las and Hanna, Emma Bou and Bressand, Florian and Lengyel, Gianna and Bour, Guillaume and Lample, Guillaume and Lavaud, Lélio Renard and Saulnier, Lucile and Lachaux, Marie-Anne and Stock, Pierre and Subramanian, Sandeep and Yang, Sophia and Antoniak, Szymon and Scao, Teven Le and Gervet, Théophile and Lavril, Thibaut and Wang, Thomas and Lacroix, Timothée and Sayed, William El},
	year = {2024},
	note = {arXiv:2401.04088 [cs]},
}

@misc{rajbhandari_deepspeed-moe_2022,
	title = {{DeepSpeed}-{MoE}: {Advancing} {Mixture}-of-{Experts} {Inference} and {Training} to {Power} {Next}-{Generation} {AI} {Scale}},
	shorttitle = {{DeepSpeed}-{MoE}},
	url = {http://arxiv.org/abs/2201.05596},
	doi = {10.48550/arXiv.2201.05596},
	urldate = {2025-04-06},
	publisher = {arXiv},
	author = {Rajbhandari, Samyam and Li, Conglong and Yao, Zhewei and Zhang, Minjia and Aminabadi, Reza Yazdani and Awan, Ammar Ahmad and Rasley, Jeff and He, Yuxiong},
	year = {2022},
	note = {arXiv:2201.05596 [cs]},
}

@article{westbrook2015cognitive,
  title={Cognitive effort: A neuroeconomic approach},
  author={Westbrook, Andrew and Braver, Todd S},
  journal={Cognitive, Affective, \& Behavioral Neuroscience},
  volume={15},
  number={2},
  pages={395--415},
  year={2015},
  publisher={Springer}
}

@article{aston2005integrative,
  title={An integrative theory of locus coeruleus-norepinephrine function: adaptive gain and optimal performance},
  author={Aston-Jones, Gary and Cohen, Jonathan D},
  journal={Annu. Rev. Neurosci.},
  volume={28},
  number={1},
  pages={403--450},
  year={2005},
  publisher={Annual Reviews}
}

@article{sara2012orienting,
  title={Orienting and reorienting: the locus coeruleus mediates cognition through arousal},
  author={Sara, Susan J and Bouret, Sebastien},
  journal={Neuron},
  volume={76},
  number={1},
  pages={130--141},
  year={2012},
  publisher={Elsevier}
}

@article{tesmer2025neurometabolic,
  title={Neurometabolic signaling and control of policy complexity},
  author={Tesmer, Alexander L and Pola, Christine Dalla and Gilli, Dino and Grujic, Nikola and Bracey, Eva F and Patriarchi, Tommaso and Peleg-Raibstein, Daria and Polania, Rafael and Burdakov, Denis},
  journal={bioRxiv},
  pages={2025--02},
  year={2025},
  publisher={Cold Spring Harbor Laboratory}
}

@article{frankle2019lotterytickethypothesisfinding,
  title={The lottery ticket hypothesis: Finding sparse, trainable neural networks},
  author={Frankle, Jonathan and Carbin, Michael},
  journal={arXiv preprint arXiv:1803.03635},
  year={2018}
}

@article{RECANATESI2022100555,
title = {A scale-dependent measure of system dimensionality},
journal = {Patterns},
volume = {3},
number = {8},
pages = {100555},
year = {2022},
issn = {2666-3899},
doi = {https://doi.org/10.1016/j.patter.2022.100555},
url = {https://www.sciencedirect.com/science/article/pii/S266638992200160X},
author = {Stefano Recanatesi and Serena Bradde and Vijay Balasubramanian and Nicholas A. Steinmetz and Eric Shea-Brown}
}

@misc{lorena2020complexclassificationproblemsurvey,
      title={How Complex is your classification problem? A survey on measuring classification complexity}, 
      author={Ana C. Lorena and Luís P. F. Garcia and Jens Lehmann and Marcilio C. P. Souto and Tin K. Ho},
      year={2020},
      eprint={1808.03591},
      archivePrefix={arXiv},
      primaryClass={cs.LG},
      url={https://arxiv.org/abs/1808.03591}, 
}

@ARTICLE{Duncan2008-kp,
  title     = "Goal neglect and Spearman's g: competing parts of a complex task",
  author    = "Duncan, John and Parr, Alice and Woolgar, Alexandra and
               Thompson, Russell and Bright, Peter and Cox, Sally and Bishop,
               Sonia and Nimmo-Smith, Ian",
  journal   = "J. Exp. Psychol. Gen.",
  publisher = "American Psychological Association (APA)",
  volume    =  137,
  number    =  1,
  pages     = "131--148",
  month     =  feb,
  year      =  2008,
  language  = "en"
}


\newpage
\appendix
\newpage
\section{Appendix}

\FloatBarrier
\subsection{Implementation Details and Code Access}
\label[appendix]{app:implementation-details}

While most details on the model implementation are described in \cref{sec:methods}, we provide some additional details here.
The GRUs in all of our layers, including routers, use the ReLU activation function and are initialized from $\mathcal{U}(-\sqrt{k},\sqrt{k})$, where $k$ is the GRU's hidden size, as is standard in PyTorch.
All models are optimized with the Schedule-Free variant of the AdamW optimizer~\cite{defazio2024road} using a learning rate of 0.01, betas of $(0.9, 0.999)$, and no weight decay.
As briefly described in \cref{sec:methods}, we use an additional embedding layer to transform the 82-dimensional one-hot encoding of the task identity into a 16-dimensional embedding vector, which is concatenated with the task stimuli before being provided to the model.
We include these details in a complete version of our training algorithm below in \cref{algo:full}, expanding on the abbreviated algorithm introduced in \cref{algo:abbreviated}.
We provide our implementation at \url{https://github.com/jackcook/mixture-of-pathways}.

\begin{algorithm}
\SetAlgoLined
\KwData{Task dataset $D$, Experts $E = \{e_1, e_2, \ldots, e_n\}$, Routers $R = \{r_1, r_2, \ldots, r_n\}$}
Initialize routers and heterogeneous experts, sampling weights and biases from $\mathcal{U}(-\sqrt{k},\sqrt{k})$\;
Process 82-dimensional one-hot task encoding with embedding layer\;
Set $h_0$ to task input, consisting of a 1-dimensional fixation input, two 16-dimensional stimuli, and a 16-dimensional task embedding at each timestep\;
\For{each training step}{
    Sample batch $b$ from $D$\;
    \For{each layer $l$}{
        Compute routing weights $w_{l,j} = \text{softmax}(r_l(h_{l-1}))$ \;
        Apply expert dropout (Not in baseline architecture; see \cref{sec:expert-dropout}) \;
        Compute expert activations $w_{l,j}$ for each expert in $l$ \;
        Combine outputs: $h_l = \sum_j w_{l,j} \cdot h_{l,j}$\;
    }
    Compute baseline model losses: $L_{\text{fix}}$ and $L_{\text{response},i}$ \; 
    Compute $L_{\text{routing}}$ loss (Not in baseline architecture; see \cref{sec:routing-stability}) \;
    Compute $L_{\text{total}} = L_{\text{fix}} + L_{\text{routing}} + \sum_i L_{\text{response},i}$\;
    Update parameters using Schedule-Free AdamW optimizer~\cite{defazio2024road}\;
}
\caption{Full Mixture-of-Pathways training protocol.}
\label{algo:full}
\end{algorithm}

\FloatBarrier
\subsection{Sample Task Visualizations and Descriptions}
\label[appendix]{app:task-visualisation}

\cref{sec:methods} briefly described the Mod-Cog task set~\cite{khona_winning_2023}. Here, we provide a more detailed description of the tasks, alongside visualizations of sample trials.

The Mod-Cog task set consists of 82 time-series-based cognitive tasks that extend the original NeuroGym framework~\cite{molano2022neurogym} through two primary modifications: integration tasks, which incorporate interval estimation based on delay periods, and sequence generation tasks, which require time-varying outputs with drifting directions. The original 20 cognitive tasks from NeuroGym serve as the foundation, with new integration-based tasks and new sequence generation tasks forming a set of 82 tasks in total. These tasks span a wide range of cognitive demands, from simple stimulus-response mappings to complex working memory and sequential decision-making challenges.

\begin{figure}
    \centering
    \includegraphics[width=\linewidth]{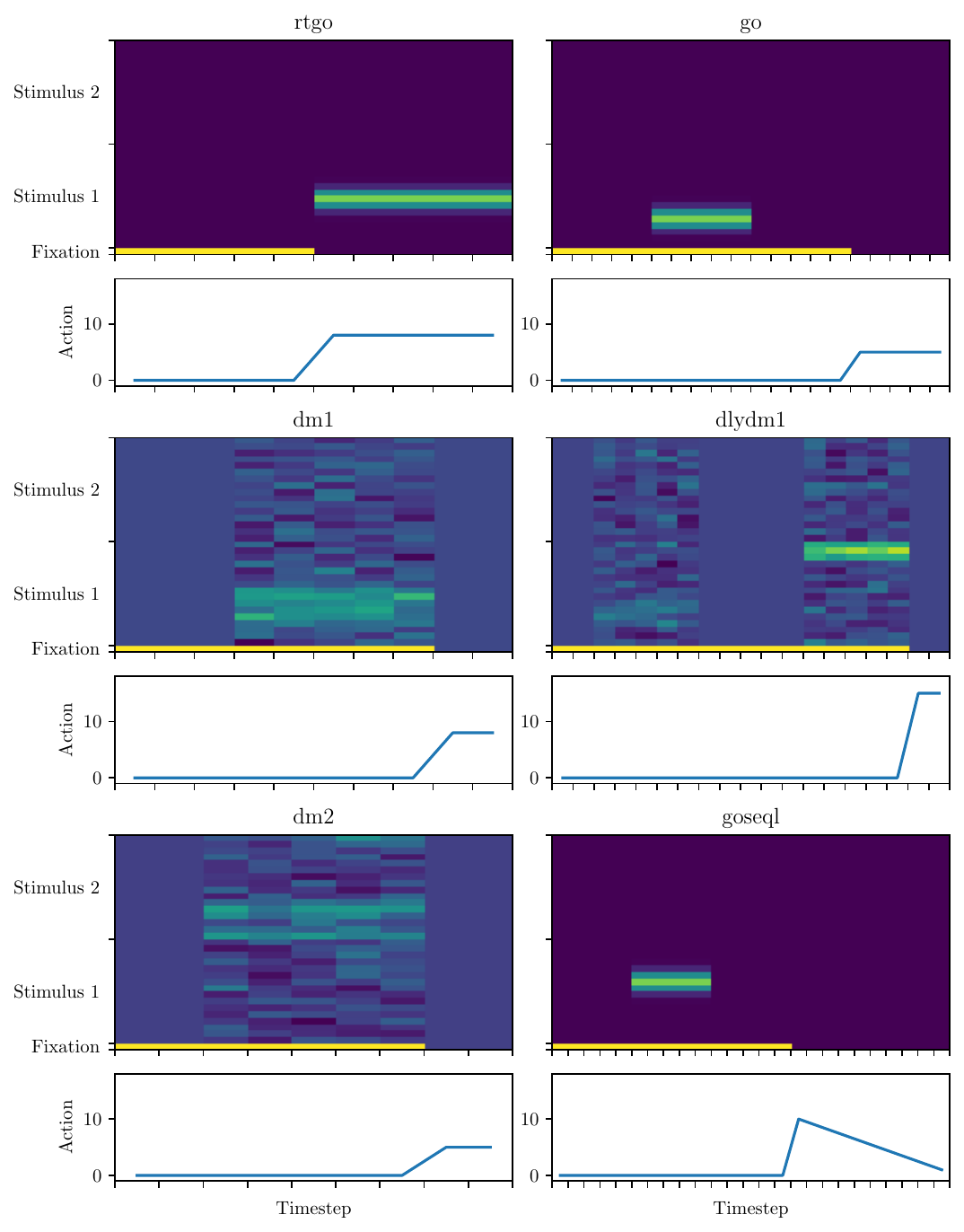}
    \caption{Model inputs and actions for the \texttt{rtgo}, \texttt{go}, \texttt{dm1}, \texttt{dlydm1}, \texttt{dm2}, and \texttt{goseqr} tasks. \cref{app:task-visualisation} provides a detailed description of each task.}
    \label{fig:task-samples}
\end{figure}

Tasks are presented as continuous time-series data, which we sample at 100-millisecond intervals. Each input consists of a 1-dimensional fixation signal, two 16-dimensional stimulus channels, and an 82-dimensional one-hot task identifier that passes through a learned embedding layer, as described in \cref{app:implementation-details}. During the fixation period of each task, models must maintain an output of zero while processing incoming stimuli. During the subsequent response period, models must return task-specific output sequences. Task difficulty varies systematically according to several factors: the complexity of decision rules (from simple detection to multi-step integration), the duration of delay periods that tax working memory, the number of stimuli that must be simultaneously tracked, and whether responses require static outputs or dynamic sequential patterns. \cref{fig:task-samples} illustrates six representative tasks that demonstrate this range of complexity.

\begin{itemize}
    \item \textbf{RTGo task (\texttt{rtgo})}: The model must immediately output the value presented in either input channel.
    \item \textbf{Go task (\texttt{go})}: The model observes two input channels and must respond with the value presented in either stimulus once the fixation period has ended.
    \item \textbf{Decision Making task (\texttt{dm1})}: The model observes brief stimulus presentations from both channels and must respond with the value in stimulus 1 that had the highest average intensity (values in stimulus 2 can be ignored). This requires integration and comparison of sensory evidence.
    \item \textbf{Delayed Decision Making task (\texttt{dlydm1})}: Similar to dm1, but includes a delay period between a decoy stimulus and the noisy stimulus, requiring working memory to maintain stimulus information.
    \item \textbf{Dual Decision Making task (\texttt{dm2})}: Similar to dm1, but the model must return the highest value from stimulus 2.
    \item \textbf{Sequential Decision Making task (\texttt{goseql})}: Based on the Go task but requiring a time-varying output that drifts in a specific direction (leftward) over the response period, combining stimulus detection with sequential motor control.
\end{itemize}

\FloatBarrier
\subsection{Loss Variants Used Across Model Types}
\label[appendix]{app:loss-overview}

In \cref{sec:pathway_formation}, we introduce architectural changes to the model that encourage pathways to form between layers. Here, we give an overview of the five different loss functions which are used to train our different model variants.

The loss function that we use to train the baseline model only includes the fixation and task response loss, as shown below in \cref{eqn:loss-baseline}.
\cref{eqn:loss-fixation} computes a mean-squared error for the fixation period where the correct output value is always zero.
This loss function is not task-specific since it can be computed across all task inputs while the fixation input is active.
\cref{eqn:loss-response} computes a task-specific cross-entropy loss between the 16 possible output values and the model's 16 output logits at each timestep.
When used to train a model, one of these response losses will be computed for each task in the set of tasks $\mathcal{T}$.

\begin{equation}
L_\text{Baseline}=L_\text{fix}+\sum_i^\mathcal{T}L_{\text{response},i}
    \label{eqn:loss-baseline}
\end{equation}

\begin{equation}
L_\text{fix}=\frac{1}{T}\sum_t^{T}\hat{y}_t^2
    \label{eqn:loss-fixation}
\end{equation}

\begin{equation}
L_{\text{response},i}=-\frac{1}{T_i} \sum_{t}^{T_i} y_t \log(\hat{y}_t)
    \label{eqn:loss-response}
\end{equation}

Here, $\hat{y}_t$ represents the model's output logits at timestep $t$ during fixation, $T$ is the total number of fixation timesteps, $y_t$ is the true target output, $\hat{y}_t$ is the model's predicted output at timestep $t$, and $T_i$ represents the total number of response timesteps for task $i$.

In \cref{sec:routing-stability}, using this baseline loss function, we find that the baseline model by itself does not converge on consistent pathways across model runs. This motivates us to create a new loss function, $L_\text{RC}$, which introduces a routing cost that penalizes the model for using more complex experts, as shown in \cref{eqn:loss-LPC}.

\begin{align}
    \label{eqn:loss-LPC}
    L_\text{RC} &= L_\text{fix}+\sum_i^\mathcal{T}(L_{\text{response},i}+\alpha LPC_i)\\
    \label{eqn:loss-LPC-2}
    LPC_i &= \frac{1}{T_i}\sum_t^{T_i}\sum_j^Ew_{i,j,t}s_j^2
\end{align}

In \cref{eqn:loss-LPC-2}, $w_{i,j,t}$ represents the routing weight assigned to expert $j$ at timestep $t$ for task $i$, $s_j$ is the size of expert $j$, and $E$ denotes the total number of experts in the model (in this manuscript, all models have three layers of three experts each, so $E=9$).
$\alpha$ is a hyperparameter that balances the trade-off between the model's performance on each task and the complexity of the experts used to solve that task.
If $\alpha$ is too large, the model will reach a local minimum where it is unable to solve any task, but it can reduce its expert usage to zero by using the skip connections in each layer and performing no computation.
On the other hand, if $\alpha$ is too small, the model will solve each task to a very high degree of accuracy, but not reduce the complexity of the experts used to solve each task, and fail to form brain-like pathways.
We found that setting $\alpha=10^{-5}$ balanced these priorities well, and used this value for all of the experiments in this work.
However, future work may investigate a better method for selecting this hyperparameter.

While the routing consistency for the model with the loss in \cref{eqn:loss-LPC} is improved as shown in \cref{fig:routing-stability}, we do find that it does not converge on a fully consistent routing pattern.
This motivates the addition of a scaling factor based on task performance, which reduces the effect of the routing loss when task performance is low. The resulting loss shown in \cref{eqn:cost-based-loss-appendix} is the final loss we use to train our \textit{Mixture-of-Pathways} model.

\begin{equation}
L=L_\text{fix}+\sum_i^\mathcal{T}(L_{\text{response},i}+\frac{\alpha LPC_i}{L_{\text{response},i}+\epsilon})
    \label{eqn:cost-based-loss-appendix}
\end{equation}

The normalization term $L_{\text{response},i}+\epsilon$ uses the task-specific response loss $L_{\text{response},i}$ to scale the routing penalty, where $\epsilon$ is a small constant added to prevent division by zero when the model achieves perfect task performance.

\FloatBarrier
\subsection{Learned Pathway Complexity and Routing Consistency}
\label[appendix]{app:penalty_calc}

\subsubsection{Calculation Example}

Here we provide a detailed example calculation of how the expert size penalty is calculated. This penalty is used within the calculation of the routing consistency described in \cref{sec:routing-stability} and also as part of the expert-usage loss from \cref{eqn:loss-LPC}.

In this equation, $E$ is the set of all experts, $s_j$ is the size of each expert, and $w_{i,j}$ is the weight assigned to expert $j$ while solving task $i$.
For example, imagine a model with three experts: a skip connection, a simple expert with 16 neurons, and a complex expert with 32 neurons.
To solve a task $i$, imagine the model sets the weight of the skip connection to $31\%$, the simple expert to $43\%$, and the complex expert to $26\%$.
The model's learned pathway complexity (LPC) for task $i$ would be 376.3, as follows:

\begin{equation}
    LPC_i=w_{i,0}s_0^2+w_{i,1}s_1^2+w_{i,2}s_2^2=(0.31)(0)^2+(0.43)(16)^2+(0.26)(32)^2=\boxed{376.3}
\end{equation}

For simplicity, this equation shows how to calculate the model's LPC at a single timestep.
To calculate the LPC for an entire task, this value should be calculated at and averaged over all of the task's timesteps.

\subsubsection{Routing Consistency Across Model Types}

\begin{figure}
    \centering
    \includegraphics[width=\linewidth]{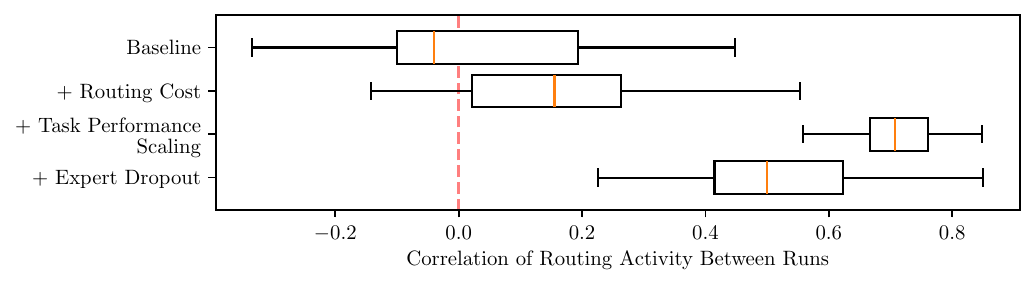}
    \caption{Routing consistency as measured by correlation of routing activity across model runs. This is a version of \cref{fig:routing-stability} which also shows the model with expert dropout. Even with expert dropout the routing consistency is significantly above the baseline model ($p<0.0001$), even though it is slightly reduced when compared to the model with routing cost and task performance scaling.}
    \label{fig:routing-stability-appendix}
\end{figure}

Using the calculated LPC values for each task, we determine the consistency of the routing decisions made by different models when solving tasks.
In \cref{fig:routing-stability-appendix}, we show a version of \cref{fig:routing-stability} with our final model, which includes expert dropout.
After the addition of expert dropout, the mean pairwise correlations of the models' routing consistency is 0.51 ($p<0.0001$). This is a reduction when compared to our model trained only with our custom routing cost and task performance scaling, however, as explained in \cref{sec:expert-dropout}, the model with the additional dropout does develop self-sufficient pathways on top of the consistency criteria, so that the model including the dropout overall better matches the pathway formation criteria. Our main investigation on what we call the \textit{Mixture-of-Pathways} model includes the expert dropout.

\FloatBarrier
\subsection{Effective Rank}
\label[appendix]{sec:effective-rank}

\begin{figure}
    \centering
    \includegraphics[width=0.75\linewidth]{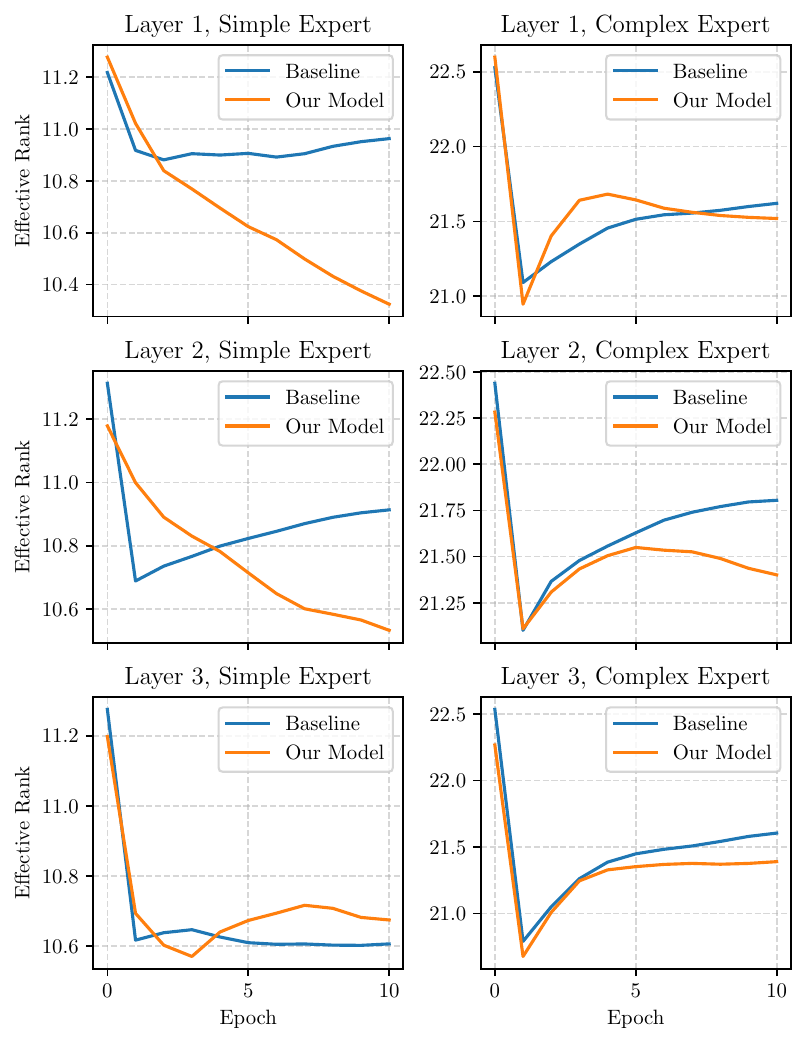}
    \caption{Changes in the effective rank of each expert's weight matrix over the course of training. Results for each configuration are averaged across 20 runs.}
    \label{fig:effective_rank}
\end{figure}

To define the $LPC$, in \cref{eqn:LPC} we use $s_j^2$ as a penalty for using expert $j$, where $s_j$ is the expert's hidden dimension, or zero in the case of a skip connection.
Intuitively, this was motivated by the $O(s_j^2)$ cost of storing each expert's weight matrix in memory, however this only roughly captures the learning capabilities of a GRU with $s_j$ neurons.
For example, as demonstrated by the lottery ticket hypothesis, it is possible that experts with large hidden dimensions may converge on low-rank solutions more easily than experts with small hidden dimensions~\cite{frankle2019lotterytickethypothesisfinding}.

To ensure that we were appropriately penalizing experts relative to each other, we analyzed the effective rank of each expert's matrix, defined as the participation ratio of the squared sum of singular values to the sum of squared singular values, $\frac{(\sum\sigma_i)^2}{\sum\sigma_i^2}$, over the course of training~\cite{RECANATESI2022100555}.
This metric measures how evenly distributed the singular values are and thus how many dimensions the matrix effectively uses.
These are shown in \cref{fig:effective_rank}.
For both models, we find that the effective rank of large experts is roughly double that of small experts, both before and after training, so that large experts always have a much larger effective rank than small experts ($p<0.0001$).
We additionally find that all experts across both our model and the baseline model decrease slightly in effective rank over training, but the differences between these models' effective ranks tends to be small and never becomes significant ($p>0.05$).
This supports the conclusion that the hidden dimension is at least a good approximation of processing complexity, but we encourage future work to consider measuring this with a scalar value.

\FloatBarrier
\subsection{Per-Task Accuracy Metrics}
\label[appendix]{app:per-task-accuracy-metrics}

In \cref{tab:per-task-accuracy-metrics} below, we report accuracy metrics on each task for three models: the baseline HMoE model described in \cref{sec:model-architecture}, our model, which includes the routing cost and task-performance scaling described in \cref{sec:routing-stability} and the expert dropout described in \cref{sec:pathway-sufficiency}, and our model without expert dropout.

\begin{longtable}[c]{lrrr}
\caption{Per-task accuracy metrics.\label{tab:per-task-accuracy-metrics}}\\
\toprule
\textbf{Task} & \textbf{Baseline} & \textbf{Our Model} & \textbf{Without Dropout} \\
\midrule
\endfirsthead
\toprule
\textbf{Task (cont.)} & \textbf{Baseline} & \textbf{Our Model} & \textbf{Without Dropout} \\
\midrule
\endhead
\bottomrule
\endfoot
\bottomrule
\endlastfoot
\textbf{Mean} & 91.1\% & 83.0\% & 90.1\% \\
\textbf{Median} & 93.3\% & 87.5\% & 91.1\% \\
\midrule
\texttt{anti} & 100.0\% & 100.0\% & 99.9\% \\
\texttt{antiseql} & 100.0\% & 99.8\% & 99.8\% \\
\texttt{antiseqr} & 100.0\% & 100.0\% & 100.0\% \\
\texttt{ctxdlydm1} & 99.5\% & 99.6\% & 99.5\% \\
\texttt{ctxdlydm1intl} & 99.8\% & 100.0\% & 99.3\% \\
\texttt{ctxdlydm1intr} & 99.6\% & 98.4\% & 99.6\% \\
\texttt{ctxdlydm1seql} & 83.5\% & 79.9\% & 84.0\% \\
\texttt{ctxdlydm1seqr} & 83.2\% & 84.1\% & 82.3\% \\
\texttt{ctxdlydm2} & 75.5\% & 71.7\% & 76.3\% \\
\texttt{ctxdlydm2intl} & 78.2\% & 70.9\% & 79.3\% \\
\texttt{ctxdlydm2intr} & 81.2\% & 82.3\% & 84.2\% \\
\texttt{ctxdlydm2seql} & 95.4\% & 90.2\% & 94.3\% \\
\texttt{ctxdlydm2seqr} & 90.7\% & 92.7\% & 90.6\% \\
\texttt{ctxdm1} & 93.0\% & 90.0\% & 93.7\% \\
\texttt{ctxdm1seql} & 93.6\% & 88.6\% & 91.0\% \\
\texttt{ctxdm1seqr} & 82.2\% & 83.0\% & 76.4\% \\
\texttt{ctxdm2} & 98.7\% & 97.9\% & 99.3\% \\
\texttt{ctxdm2seql} & 99.5\% & 97.5\% & 98.5\% \\
\texttt{ctxdm2seqr} & 99.4\% & 92.1\% & 99.0\% \\
\texttt{dlyanti} & 99.7\% & 98.4\% & 99.5\% \\
\texttt{dlyantiintl} & 99.8\% & 39.1\% & 98.7\% \\
\texttt{dlyantiintr} & 99.6\% & 44.3\% & 94.1\% \\
\texttt{dlyantiseql} & 96.2\% & 38.4\% & 98.4\% \\
\texttt{dlyantiseqr} & 98.7\% & 37.6\% & 93.4\% \\
\texttt{dlydm1} & 91.8\% & 87.5\% & 88.6\% \\
\texttt{dlydm1intl} & 93.0\% & 88.5\% & 91.8\% \\
\texttt{dlydm1intr} & 91.7\% & 89.2\% & 88.7\% \\
\texttt{dlydm1seql} & 92.8\% & 86.0\% & 89.3\% \\
\texttt{dlydm1seqr} & 90.5\% & 90.6\% & 88.8\% \\
\texttt{dlydm2} & 90.2\% & 87.0\% & 90.5\% \\
\texttt{dlydm2intl} & 92.2\% & 84.1\% & 86.8\% \\
\texttt{dlydm2intr} & 92.1\% & 85.1\% & 88.1\% \\
\texttt{dlydm2seql} & 82.4\% & 78.0\% & 74.3\% \\
\texttt{dlydm2seqr} & 82.1\% & 78.9\% & 76.8\% \\
\texttt{dlygo} & 80.2\% & 59.9\% & 76.0\% \\
\texttt{dlygointl} & 78.5\% & 60.8\% & 79.6\% \\
\texttt{dlygointr} & 82.6\% & 61.5\% & 84.3\% \\
\texttt{dlygoseql} & 85.5\% & 60.4\% & 82.0\% \\
\texttt{dlygoseqr} & 74.3\% & 62.6\% & 76.8\% \\
\texttt{dm1} & 76.3\% & 61.7\% & 75.0\% \\
\texttt{dm1seql} & 78.8\% & 57.3\% & 78.4\% \\
\texttt{dm1seqr} & 81.0\% & 63.4\% & 76.6\% \\
\texttt{dm2} & 100.0\% & 98.0\% & 99.9\% \\
\texttt{dm2seql} & 99.7\% & 97.6\% & 99.2\% \\
\texttt{dm2seqr} & 100.0\% & 98.0\% & 99.9\% \\
\texttt{dmc} & 99.9\% & 95.5\% & 99.8\% \\
\texttt{dmcintl} & 99.8\% & 99.0\% & 99.4\% \\
\texttt{dmcintr} & 99.5\% & 98.0\% & 99.8\% \\
\texttt{dmcseql} & 81.9\% & 75.2\% & 81.1\% \\
\texttt{dmcseqr} & 81.5\% & 75.0\% & 81.7\% \\
\texttt{dms} & 75.2\% & 66.5\% & 77.2\% \\
\texttt{dmsintl} & 73.4\% & 64.8\% & 77.3\% \\
\texttt{dmsintr} & 79.7\% & 72.8\% & 86.0\% \\
\texttt{dmsseql} & 95.1\% & 87.6\% & 92.8\% \\
\texttt{dmsseqr} & 95.2\% & 88.6\% & 90.6\% \\
\texttt{dnmc} & 93.2\% & 87.3\% & 88.7\% \\
\texttt{dnmcintl} & 93.1\% & 86.2\% & 89.7\% \\
\texttt{dnmcintr} & 80.8\% & 79.0\% & 77.1\% \\
\texttt{dnmcseql} & 98.5\% & 93.0\% & 98.9\% \\
\texttt{dnmcseqr} & 99.7\% & 94.3\% & 98.3\% \\
\texttt{dnms} & 98.8\% & 91.3\% & 98.8\% \\
\texttt{dnmsintl} & 99.3\% & 93.6\% & 98.1\% \\
\texttt{dnmsintr} & 99.9\% & 99.4\% & 100.0\% \\
\texttt{dnmsseql} & 100.0\% & 97.4\% & 99.6\% \\
\texttt{dnmsseqr} & 100.0\% & 99.1\% & 99.4\% \\
\texttt{go} & 100.0\% & 94.9\% & 99.7\% \\
\texttt{goseql} & 99.9\% & 96.7\% & 99.4\% \\
\texttt{goseqr} & 100.0\% & 94.0\% & 99.7\% \\
\texttt{multidlydm} & 81.8\% & 74.5\% & 81.9\% \\
\texttt{multidlydmintl} & 78.1\% & 77.3\% & 80.0\% \\
\texttt{multidlydmintr} & 74.1\% & 66.7\% & 76.0\% \\
\texttt{multidlydmseql} & 73.1\% & 66.6\% & 71.8\% \\
\texttt{multidlydmseqr} & 79.7\% & 71.6\% & 79.4\% \\
\texttt{multidm} & 94.0\% & 90.0\% & 90.7\% \\
\texttt{multidmseql} & 94.3\% & 88.9\% & 93.1\% \\
\texttt{multidmseqr} & 93.3\% & 85.9\% & 92.6\% \\
\texttt{rtanti} & 92.6\% & 86.1\% & 91.2\% \\
\texttt{rtantiseql} & 82.2\% & 79.5\% & 81.6\% \\
\texttt{rtantiseqr} & 99.7\% & 95.2\% & 98.9\% \\
\texttt{rtgo} & 98.7\% & 94.9\% & 98.9\% \\
\texttt{rtgoseql} & 99.3\% & 91.9\% & 98.0\% \\
\texttt{rtgoseqr} & 98.1\% & 93.1\% & 99.0\% \\

\end{longtable}

\FloatBarrier
\subsection{Task-Specific Expert Usage Patterns}
\label[appendix]{app:task-specific-expert-usage}

\begin{figure}
    \centering
    \includegraphics[width=\linewidth]{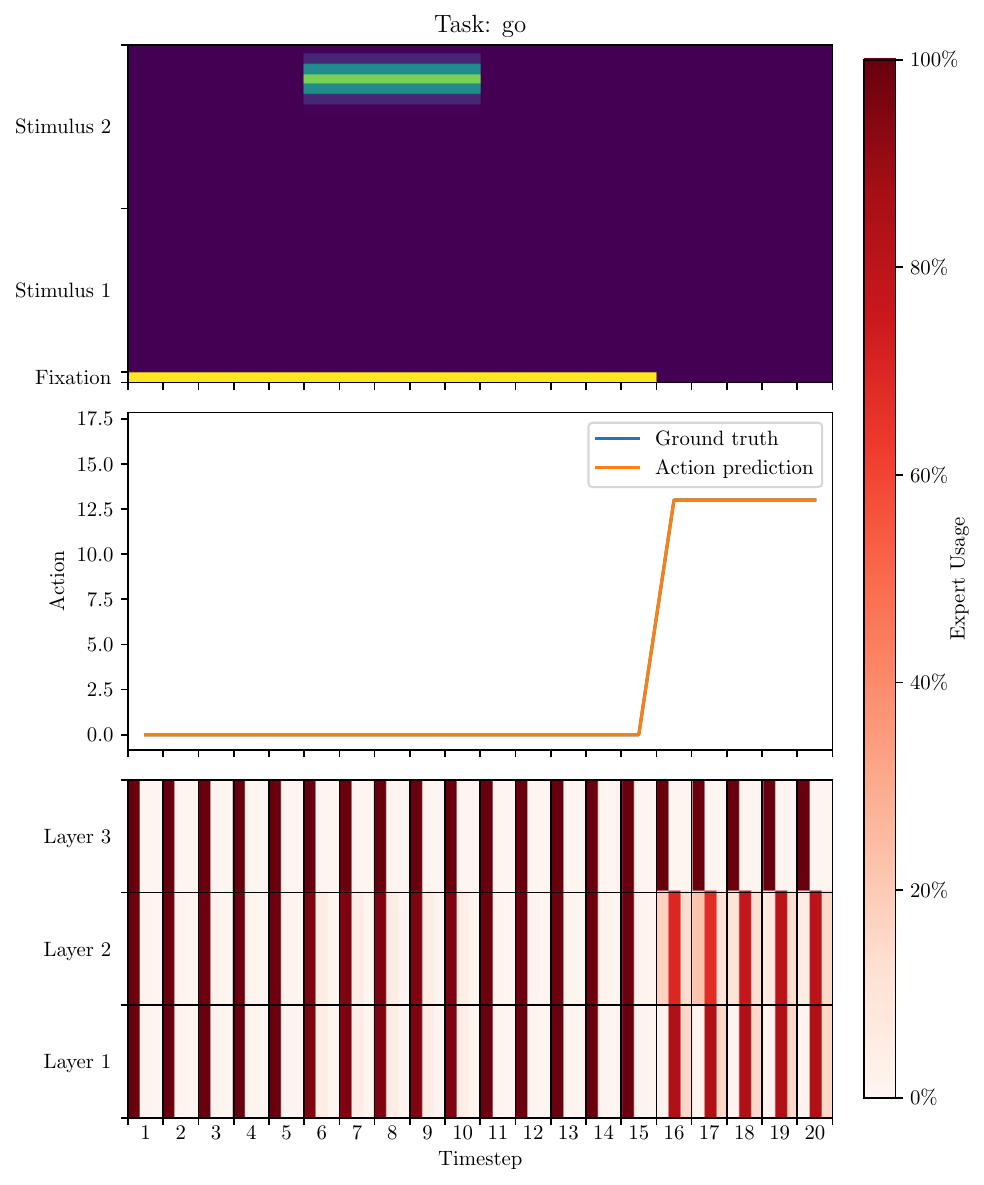}
    \caption{In the go task we see that models rely on extremely simple pathways for their decision making until activating model complex pathways during the response period.}
    \label{fig:trial-go}
\end{figure}

\begin{figure}
    \centering
    \includegraphics[width=\linewidth]{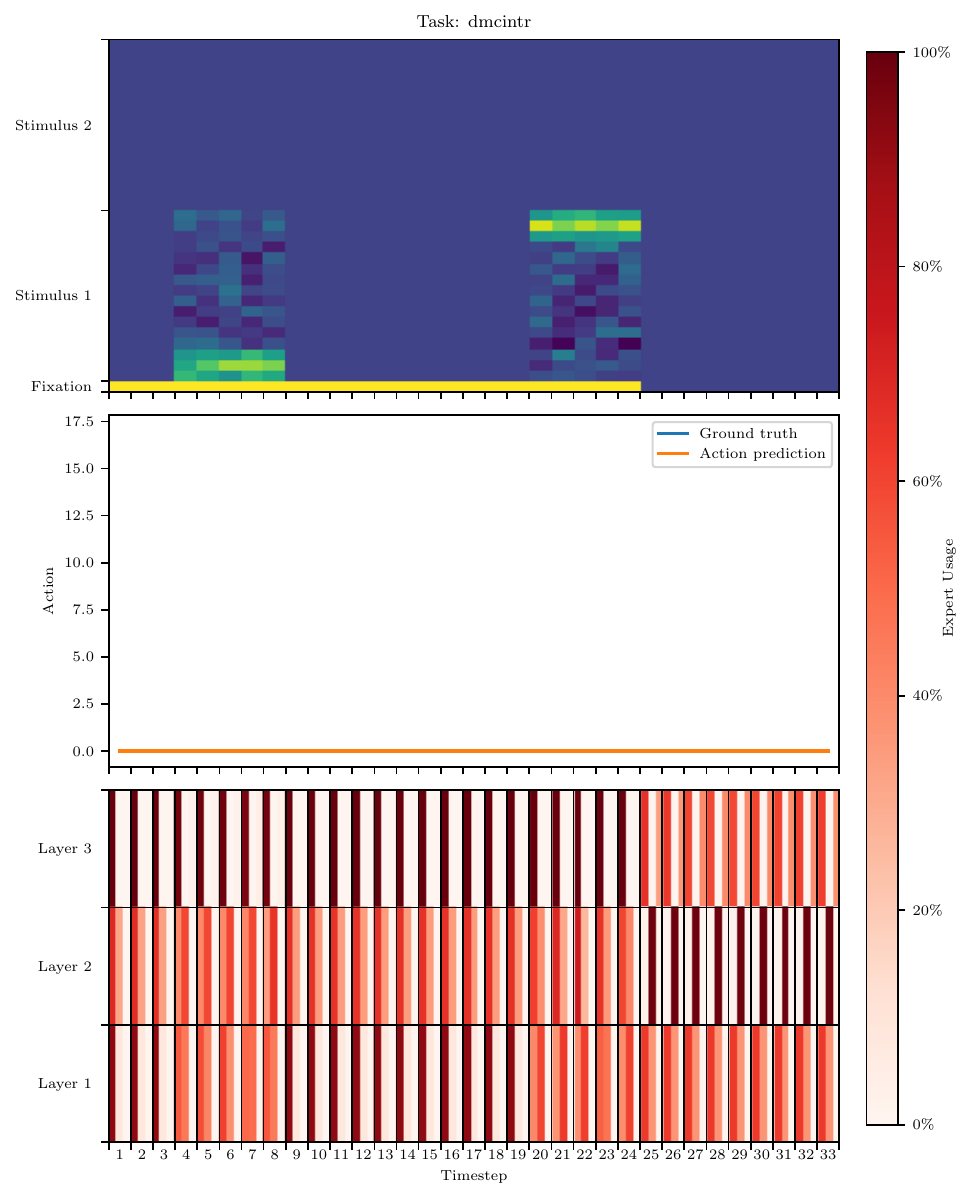}
    \caption{During more complex tasks such as dmcintr we see that models us a a very dynamic and rich set of pathways throughout the duration of a trial to return the correct response.}
    \label{fig:trial-dmcintr}
\end{figure}

In \cref{fig:trial-go,fig:trial-dmcintr}, we show two samples of tasks completed by our model.
At each timestep, we plot the model's decisions in orange, which overlap with the blue ground truth values, indicating that in these trials, the model always returned the correct answer.
We additionally plot the expert usage of the model at each timestep.
At each timestep, three bars are shown for each layer, with their color indicating the usage of the skip connection, the simple expert, and the complex expert, in that order.
In the \texttt{go} trial shown in \cref{fig:trial-go}, the model primarily uses the skip connections at each timestep during the fixation period, in which it only needs to return zero.
During the response period, the model moves its computations toward a more complex pathway, primarily using the simple experts in layers 1 and 2, and the skip connection in layer 3.
In the more complicated \texttt{dmcintr} trial shown in \cref{fig:trial-dmcintr}, the model switches between pathways multiple times depending on its needs, which vary between working memory and computation. This shows that our model has formed distinct pathway and modes of processing information which it dynamically switches between. As a result, the model opens up the possibility of analyzing the detailed dynamics of how pathways are combined over tasks with time courses that include varying computational demands, to learn which principles underlie such coordination processes.

\FloatBarrier
\subsection{Unclustered Expert Usage Patterns Across Tasks}
\label[appendix]{app:expert-usage}

\begin{figure}[hbt!]
    \centering
    \includegraphics[width=\linewidth]{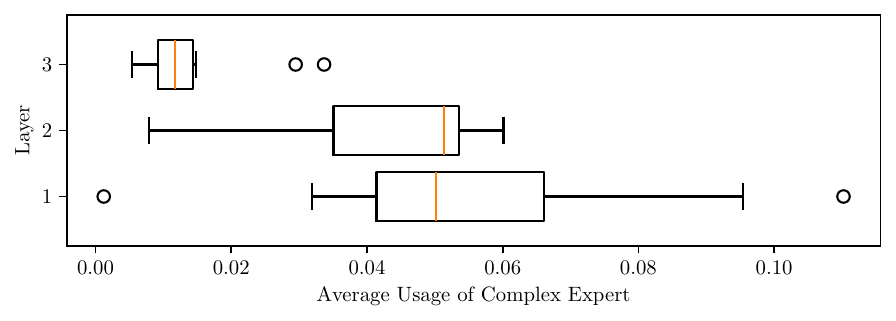}
    \caption{Average usage of the most complex experts for each cluster derived in \cref{fig:clustered-complex-expert}, split by layer index.}
    \label{fig:clustered-complex-expert}
\end{figure}

\begin{figure}
    \centering
    \includegraphics[width=\linewidth]{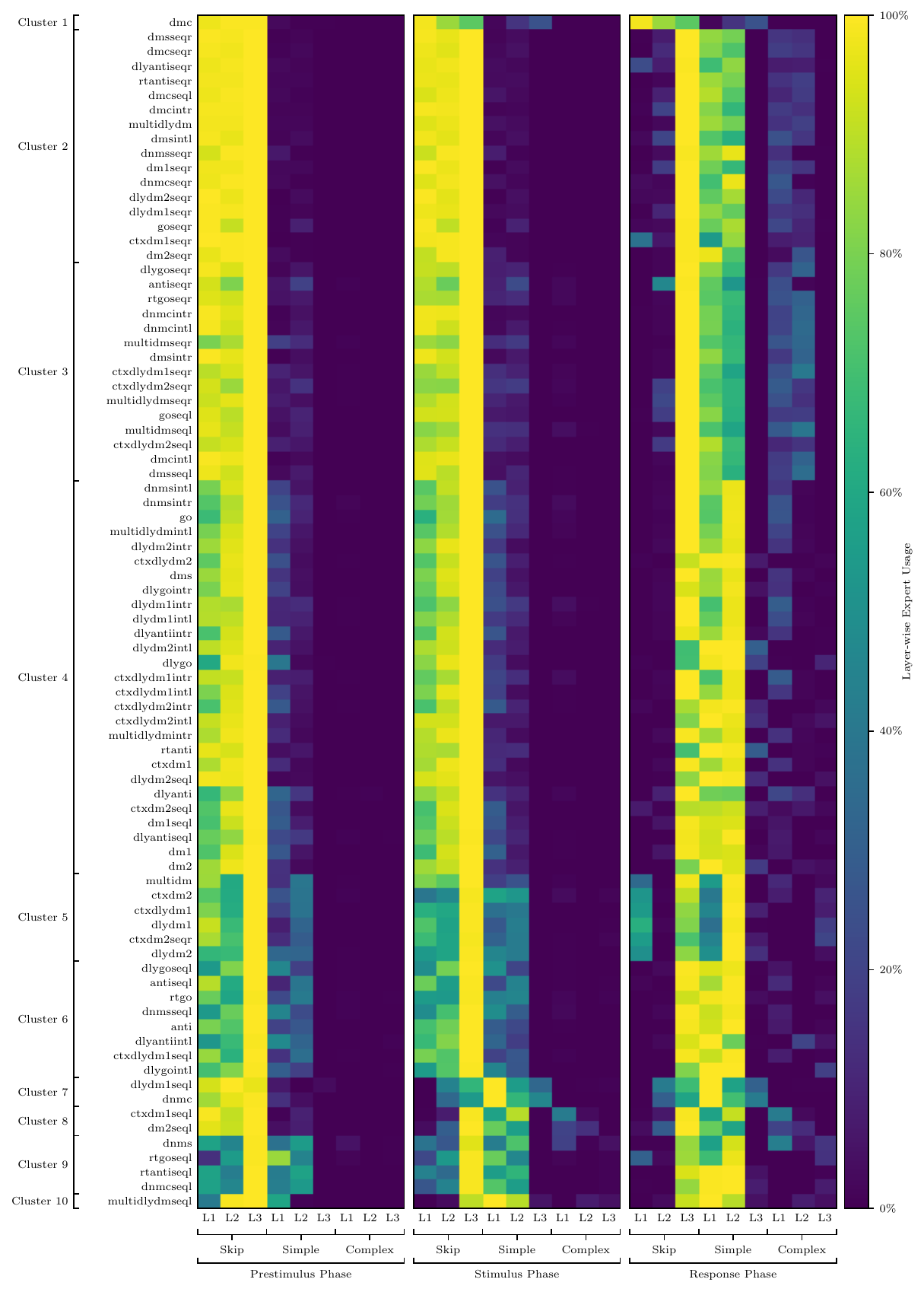}
    \caption{Layer-wise expert usage averaged over three phases of each task: the pre-stimulus phase, during which the task is known but no input data has been given to the model, the stimulus phase, during which input data is being given to the model, and the response phase, during which the model needs to calculate and return the correct response. In each phase, expert usages sum to 100\% within each layer, i.e. usages of the skip connection, simple expert, and complex expert of layer 1, denoted by ``L1'', add up to 100\%. Tasks are grouped into 10 clusters, shown along the left, based on similarities in their routing patterns. A contrasting version of this figure for a baseline model trained without our cost-based loss and dropout is shown in \cref{fig:expert-usages-full-standard}.}
    \label{fig:expert-usages-full}
\end{figure}

\begin{figure}
    \centering
    \includegraphics[width=\linewidth]{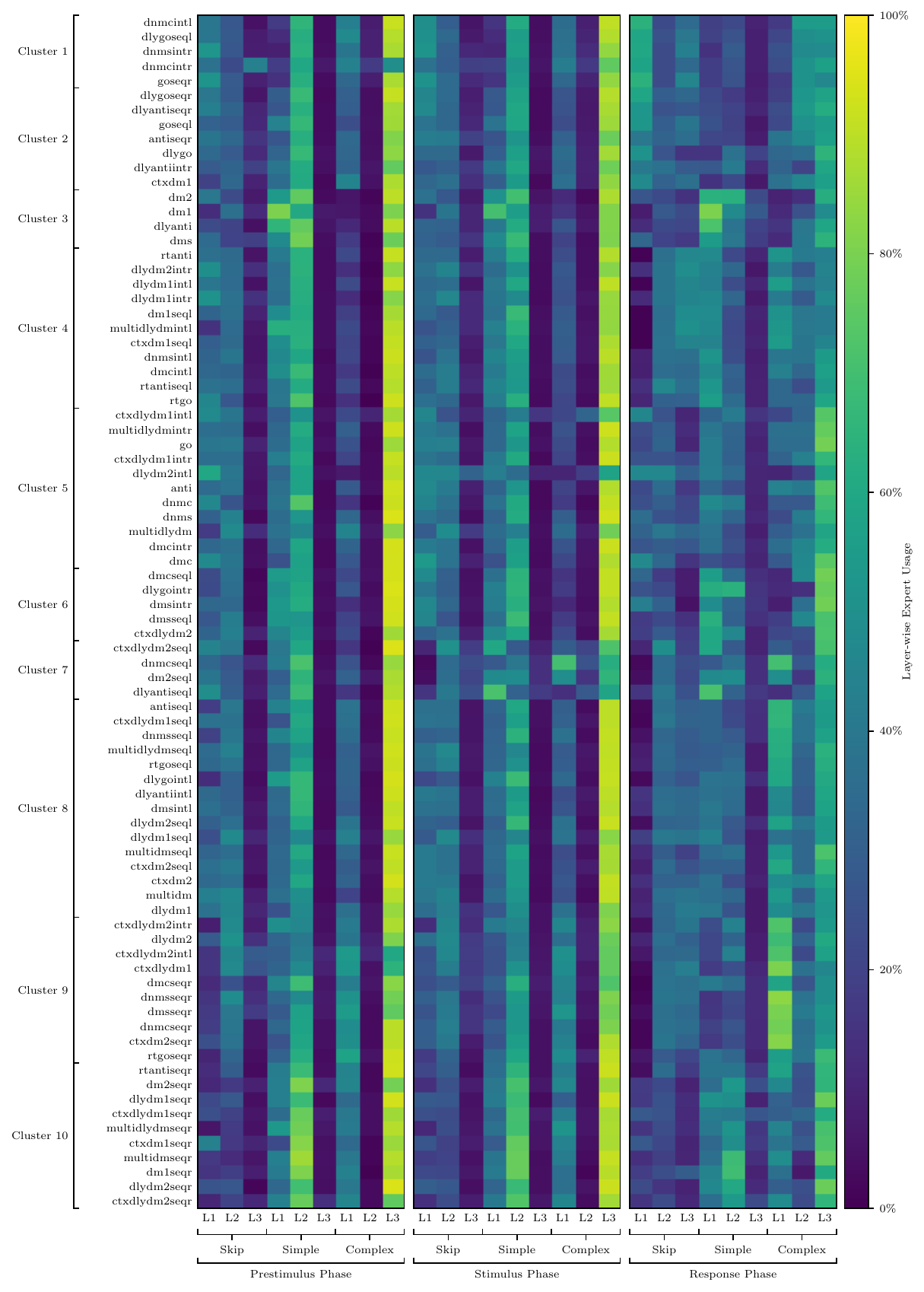}
    \caption{A separate version of \cref{fig:expert-usages-full} for a model trained without our custom routing cost, loss normalization, and expert dropout. The pathways that form are much less distinct, and are also much less stable.}
    \label{fig:expert-usages-full-standard}
\end{figure}

In \cref{fig:expert-usages-full,fig:expert-usages-full-standard}, we show the unclustered expert usages during three phases of each task.
Each task has a different number of timesteps, so in order to condense expert usage into a single figure, we averaged expert usage over three phases which are shared by each task: a pre-stimulus phase, during which the task identity is known but input data has not yet been presented, a stimulus phase, during which the model is observing input data, and a response phase, during which the model needs to output its responses.
In these figures, tasks are sorted based on the same clusters shown in \cref{fig:expert-usages-clustered} for visual clarity.
Notably, the complex expert is rarely used in the first two phases, during which the model outputs zero, but is commonly used during the response phase of complex tasks. When analyzing the average usage of the most complex experts over the clusters shown in \cref{fig:expert-usages-clustered}, we find that the cluster with the highest reliance of the most complex experts uses those with an average routing weight of 0.11. In \cref{fig:clustered-complex-expert} we show the distribution of complex expert usage by layers of the model, over clusters derived in \cref{fig:expert-usages-clustered}, meaning each data points here is one of the 10 clusters.

\FloatBarrier
\subsection{Task Difficulty}
\label[appendix]{app:task-difficulty}

\subsubsection{Number of Training Steps}
\label[appendix]{app:difficulty_numSteps}

\begin{figure}
    \centering
    \includegraphics[width=\linewidth]{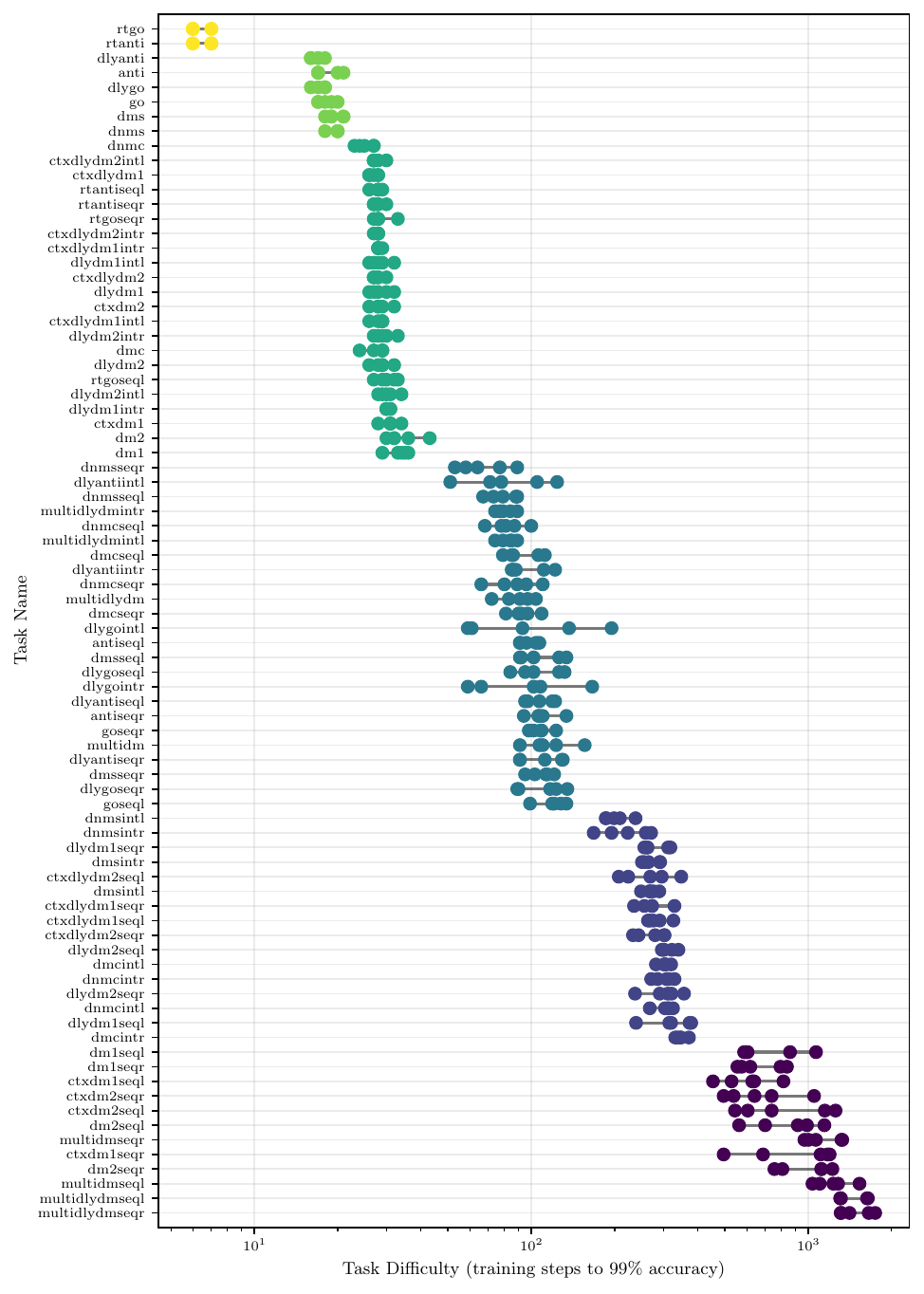}
    \caption{Five measurements of task difficulty made for each task in the Mod-Cog task suite~\cite{khona_winning_2023}. Tasks are sorted by the median of the five measurements. Interestingly, the tasks form groups around similar difficulty levels, which we indicate with the color of each dot.}
    \label{fig:task-difficulty}
\end{figure}

To measure each task's difficulty, we train five recurrent neural networks (GRUs), each with 64 neurons, and record how many training steps it takes each model to solve that task to 99\% accuracy.
The task's difficulty is then reported as the median number of steps from these five training runs.
\cref{fig:task-difficulty} shows these measurements for all 82 tasks in the Mod-Cog task suite~\cite{khona_winning_2023}.

\subsubsection{Number of Rules}

\begin{figure}
    \centering
    \includegraphics[width=\linewidth]{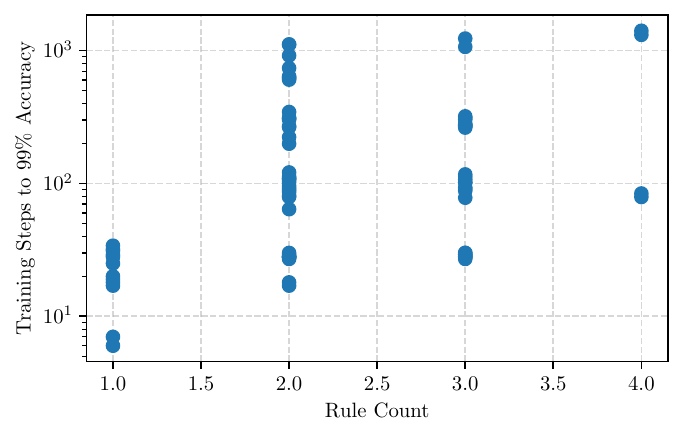}
    \caption{The number of rules that form a Mod-Cog task is correlated with the number of steps it takes a single RNN to learn the task ($r=0.39$, $p<0.0001$).}
    \label{fig:task-difficulty-metrics}
\end{figure}

There are many ways to characterize the difficulty of learning problems~\cite{lorena2020complexclassificationproblemsurvey}, but no universal complexity measure has been developed to date.
We believe ``training steps needed to learn the task,'' as discussed above in \cref{app:difficulty_numSteps}, is a sensible measure because this can be measured without any implicit biases, and takes into account task demands such as working memory and any other factors which make inference challenging~\cite{lorena2020complexclassificationproblemsurvey}.
At the same time, it remains unclear whether such a complexity measure would neatly map onto what humans or animals perceive to be ``difficult tasks,'' which is often linked to the number of rules in a task~\cite{Duncan2008-kp}.
However, this can be easily tested, as Mod-Cog tasks are created based on combinations of different motifs and rules.
For example, \cref{fig:task-samples} shows that the ``Delayed Decision Making'' task (\texttt{dlydm1}) is an altered version of the standard ``Decision Making'' task (\texttt{dm1}) with the added ``Delay'' rule (\texttt{dly}).

We find that our difficulty metric is in fact correlated with the number of rules in each task ($r=0.39$; $p<0.0001$, shown in \cref{fig:task-difficulty-metrics}), and that our model exhibits an even stronger correlation for the brain-like finding in \cref{fig:task-complexities} when this is used as the difficulty metric ($r=0.57$, $p<0.0001$) compared to the baseline model ($r=-0.09$, $p=0.4288$). At the same time, using ``number of rules'' as the actual difficulty metric has two downsides: (a) it is a discrete and ordinal measurement from 1 to 4 with less statistical power, and (b) some rules are harder to learn than others (i.e. \texttt{go} vs. \texttt{dm1}). This suggests to us that our current convergence-based complexity measure is, at least in this specific task environment, a difficulty metric providing a better link to both the brain and GRU-based machine learning models.
\end{document}